\documentclass[nolongbibliography,noeprint,notitlepage,superscriptaddress,reprint,prd,nofootinbib,floatfix,]{revtex4-2}

\usepackage{graphicx}
\usepackage{bm}
\usepackage{amsmath}
\usepackage{dcolumn}
\usepackage[flushleft]{threeparttable}
\usepackage{makecell}
\usepackage{booktabs}
\usepackage{comment}

\usepackage[citecolor=blue,hypertexnames=false]{hyperref}
\hypersetup{
    colorlinks=true,       
    linkcolor=blue,                 
    filecolor=blue,      
    urlcolor=blue
}

\usepackage[pagewise,left]{lineno}

\graphicspath{{Images/}}

\newcommand{\hi}{H~{\sc i}}

\newcommand{\nodata}{\centering\arraybackslash --}

\begin{document}

\preprint{Prepared for submission to PRD}

\title{Dark Matter Interpretation of the \textit{Fermi}-LAT Observations \\ Toward the Outer Halo of M31}

\author{Christopher M. Karwin}
\email{ckarwin@clemson.edu}
\affiliation{Department of Physics and Astronomy, Clemson University, Clemson, SC, USA}
\affiliation{Department of Physics and Astronomy, University of California, Irvine, CA, USA}

\author{Simona Murgia}
\email{smurgia@uci.edu}
\affiliation{Department of Physics and Astronomy, University of California, Irvine, CA, USA}

\author{Igor V. Moskalenko}
\email{imos@stanford.edu}
\affiliation{Hansen Experimental Physics Laboratory and Kavli Institute for Particle Astrophysics and Cosmology, Stanford University, Stanford, CA, USA}

\author{Sean P. Fillingham}
\affiliation{Department of Astronomy, University of Washington, Seattle, WA, USA}
\affiliation{Department of Physics and Astronomy, University of California, Irvine, CA, USA}

\author{Anne-Katherine Burns}
\affiliation{Department of Physics and Astronomy, University of California, Irvine, CA, USA}

\author{Max Fieg}
\affiliation{Department of Physics and Astronomy, University of California, Irvine, CA, USA}

\date{\today}

\begin{abstract}
An excess $\gamma$-ray signal toward the outer halo of M31 has recently been reported. Although other explanations are plausible, the possibility that it arises from dark matter (DM) is valid. In this work we interpret the excess in the framework of DM annihilation, using as our representative case WIMP DM annihilating to bottom quarks, and we perform a detailed study of the systematic uncertainty in the $J$-factor for the M31 field. We find that the signal favors a DM particle with a mass of $\sim$46--73 GeV. While the mass is well constrained, the systematic uncertainty in the cross-section spans 2.5 orders of magnitude, ranging from $\sim$8$\times 10^{-27}-4 \times 10^{-24} \ \mathrm{cm^3 \ s^{-1}}$. This high uncertainty is due to two main factors, namely, an uncertainty in the substructure nature and geometry of the DM halos for both M31 and the Milky Way (MW), and correspondingly, an uncertainty in the contribution to the signal from the MW's DM halo along the line of sight. However, under the conditions that the minimum subhalo mass is $\lesssim 10^{-6} \ M_\odot$ and the actual contribution from the MW's DM halo along the line of sight is at least $\sim$30$\%$ of its total value, we show that there is a large overlap with the DM interpretations of both the Galactic center (GC) excess and the antiproton excess, while also being compatible with the limits for the MW dwarf spheroidals. More generally, we summarize the results from numerous complementary DM searches in the energy range 10 GeV $-$ 300 GeV corresponding to the GC excess and identify a region in parameter space that still remains viable for discovery of the DM particle.   
\end{abstract}

\maketitle

\section{Introduction}

Observational evidence for dark matter (DM) in M31 comes from measurements of its rotational velocity curve~\cite{babcock1939rotation,Rubin:1970zza,roberts1975rotation,carignan2006extended,2010A&A...511A..89C}. These observations provide coarse-grained properties of the DM distribution near the central regions of the halo where the galaxy resides. With the existing data, the fine-grained structure of DM and its distribution outside of the galaxy is primarily inferred from simulated halos. Within the standard cosmological paradigm, M31's DM halo is expected to extend well beyond the galactic disk, and it is also expected to contain a large amount of substructure. However, there is currently a high level of uncertainty regarding the exact nature of the halo properties, i.e. the geometry, extent, and substructure content, especially on galactic scales~\cite{Kamionkowski:1997xg,Braun:1998ik,blitz1999high,Bullock:1999he,deHeij:2002ne,Braun:2003ey,Helmi:2003pp,Bailin:2004wu,Allgood:2005eu,Bett:2006zy,Hayashi:2006es,Kuhlen:2007ku,Banerjee:2008kt,Springel:2008cc,diemand2008clumps,Zemp:2008gw,Saha:2009dt,Law:2009yq,Banerjee:2011rr,Gao:2011rf,Ishiyama:2014uoa, garrison2014elvis,Velliscig:2015ffa,Bernal:2016guq,Moline:2016pbm,garrison2017not,karwin2019fermi}.

Due to its mass and proximity, the detection sensitivity of M31 to DM searches with $\gamma$-rays is competitive with the Milky Way (MW) dwarf spheroidal galaxies, particularly if the signal is sufficiently boosted by substructures~\cite{Falvard:2002ny,Fornengo:2004kj,Mack:2008wu,Dugger:2010ys,Conrad:2015bsa,Gaskins:2016cha}. Moreover, M31 is predicted to be the brightest extragalactic source of DM annihilation~\cite{lisanti2018search,lisanti2018mapping}.

A detailed study of the $\gamma$-ray emission observed towards M31's outer halo has recently been made in Ref.~\cite{karwin2019fermi}. In that study evidence is found for an excess signal that appears to be distinct from the conventional MW foreground, having a total radial extension upwards of $\sim$120--200 kpc from the center of M31. One possible explanation for the signal is that it arises from cosmic rays (CRs) which have escaped the galactic disk and are interacting with the gas of M31's circumgalactic medium. However, the spectral properties of the observed emission do not seem to be consistent with standard CR scenarios~\cite{karwin2019fermi}. The other main physical interpretation is that the signal arises from DM, which is thought to be the dominant component in the outer regions of the galaxy. 

$\gamma$-ray emission from M31's inner galaxy has also been detected, but the exact nature of the emission still remains an open question, as the morphology of the signal doesn't appear to trace regions rich in gas and star formation~\cite{Fermi-LAT:2010kib,Pshirkov:2016qhu,Ackermann:2017nya,eckner2018millisecond,mcdaniel2018multiwavelength,karwin2019fermi,DiMauro:2019frs,McDaniel:2019niq}. On the other hand, the total $\gamma$-ray luminosity is found to be in general agreement with the well-known scaling relationship between the $\gamma$-ray luminosity and infrared  luminosity (8--1000 $\mu$m) for star-forming galaxies~\cite{Ajello:2020zna}. Ultimately, a better determination of the $\gamma$-ray signal from M31's inner region is still needed, which will require a refinement of the underlying gas maps (\hi) used to model the Galactic foreground emission, as the current maps may be holding a fraction of gas that actually resides in the M31 system~\cite{karwin2019fermi}. The Doppler-shifted velocity of the gas, together with the Galactic rotation curve, is used to separate the MW and M31 gas. The uncertainty arises from two main conditions. First, there is a partial overlap of the rotational velocities for M31 and the MW. Second, M31 is at a fairly high latitude where there is an increased uncertainty in the rotational speed of the MW gas, which is measured in the Galactic disk.   

In this work we interpret the excess $\gamma$-ray emission observed towards M31's outer halo in the framework of DM annihilation. We consider WIMP (i.e. weakly interacting massive particle) DM, and focus the analysis on the uncertainties associated with the properties of the DM halo. Moreover, we consider a realistic observational perspective, in which the line of sight towards M31's outer DM halo naturally extends through a similar DM halo around the MW. In general, this is not directly accounted for when modeling the MW foreground $\gamma$-ray emission, and can significantly impact the results. 

The paper is organized as follows. In Section~\ref{sec:M31_outer_halo} we give a qualitative description of M31's outer halo. In Section~\ref{sec:analysis} we present the M31 data, DM fit, and analytical $J$-factor calculations. In Section~\ref{sec:DM_parameterspace} we present results for our best-fit models, and we consider these results in the context of the Galactic center (GC) excess, and more generally, in the context of the current status of DM indirect detection. In Section~\ref{sec:conclusion} we conclude. Additional details for the complementary DM searches we consider are given in Appendix~\ref{sec:DM_parameter_space}.

\section{M31's Outer Halo}
\label{sec:M31_outer_halo}
For observations of $\gamma$-ray emission arising from DM annihilation towards M31's outer halo, the total signal would ostensibly contain emission from the MW's DM halo along the line of sight, emission from the local filamentary structure connecting the MW and M31, and emission from the entire DM halo of M31, plus any secondary emission (from  M31 and the MW). For the MW halo, a DM signal should be pretty bright, but since the observation occurs from within the halo, the emission can be easily confused with the isotropic component (and other components of the MW interstellar emission model (IEM)). For M31, we observe the entire halo from the outside, and therefore we see the total integral signal. Thus M31 is advantageous for halo searches with $\gamma$-rays because it breaks the observational degeneracy.

\begin{figure}[t]
\centering
\includegraphics[width=0.49\textwidth]{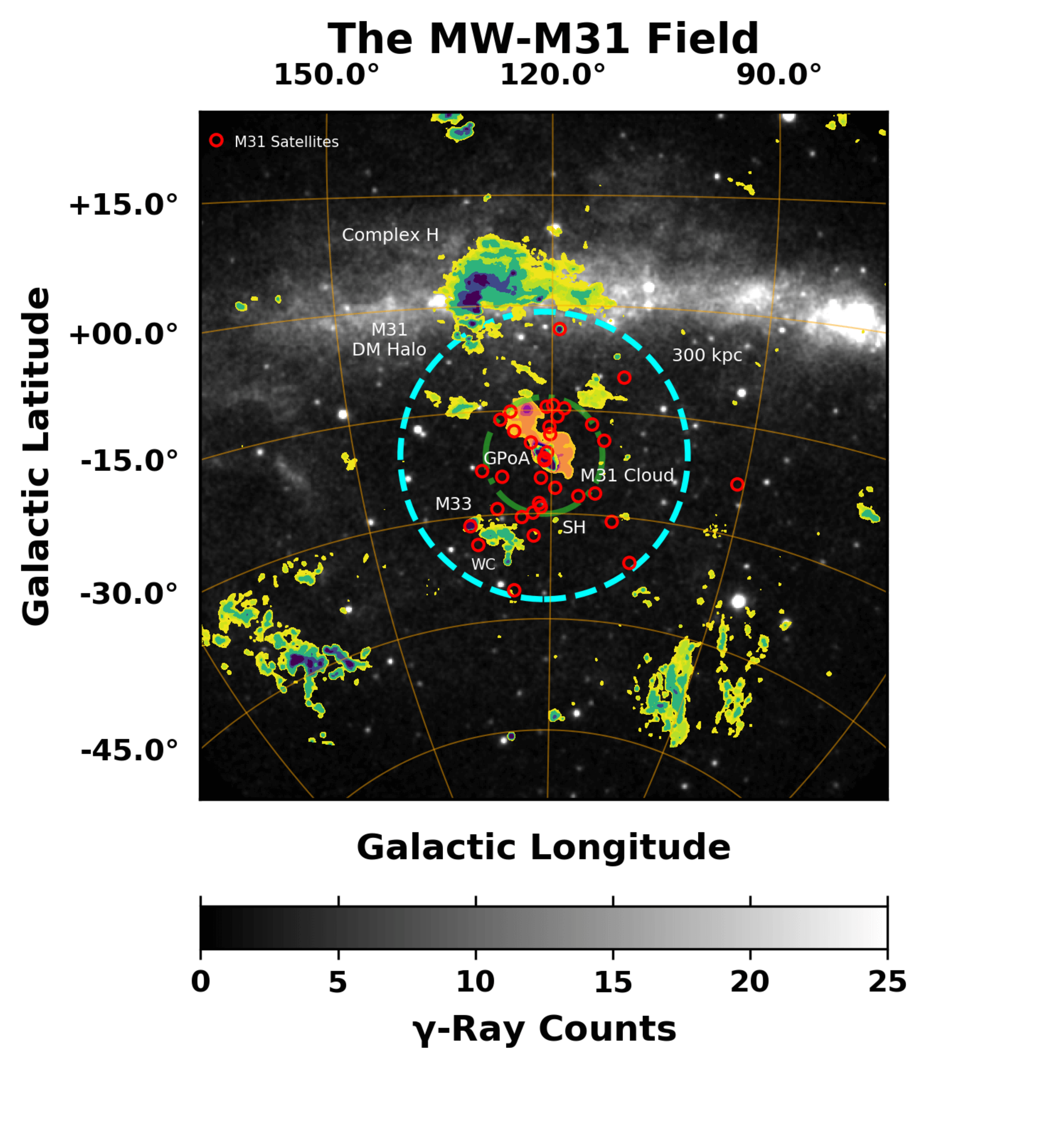}
\caption[The MW-M31 Field]{The line of sight looking towards M31's outer halo. The size of M31's DM halo is indicated with a dashed cyan circle, which corresponds to a projected radius of 300 kpc, for an M31-MW distance of 785 kpc. The dash-dot lime-green circle shows the outer boundary of the SH region ($r_{\mathrm{tan}}=117$ kpc), which we use for the DM fit. M31's population of satellite galaxies is shown with red open circles. M33 can be seen in the lower left corner. Also plotted are some notable gas clouds in the region, namely, the M31 cloud (orange region surrounding the M31 disk), Wright's cloud (WC), and Complex H. See text for more details.}
\label{fig:M31_DM_View}

\end{figure}

Figure~\ref{fig:M31_DM_View} provides a qualitative description of M31's outer halo, including an accounting of some notable structures along the line of sight that may provide hints of the DM distribution. The $\gamma$-ray counts map (shown in black and white) is from Ref.~\cite{karwin2019fermi}. The bright emission along zero degree latitude is the plane of the MW. The size of M31's DM halo is indicated with a dashed cyan circle, which corresponds to a projected radius of 300 kpc, for an M31-MW distance of 785 kpc. The dash-dot lime-green circle shows the outer boundary of the spherical halo (SH) region, which we use for the DM fit, as discussed in Section~\ref{sec:analysis}. M31's satellite population is shown with open red circles. A subset of the satellites in M31 (which are thought to reside within DM substructures) are known to be positioned within a large thin plane (The Great Plane of Andromeda, GPoA); and likewise, a subset of the MW satellites are known to be part of a large planar structure as well (The Vast Polar Structure of the Milky Way)~\cite{Kroupa:2004pt,Pawlowski:2012vz,Conn:2013iu,Ibata:2013rh,Pawlowski:2013kpa,Hammer:2013bga,Pawlowski:2018sys}. In addition, the satellite system of M31 is highly lopsided, as about 80\% of its satellites lie on the side closest to the MW~\cite{Conn:2013iu,pawlowski2017lopsidedness}. For members of the GPoA, those to the north of M31 recede from us, and those to the south of M31 move toward us, in the plane of rotation. 

Also shown in Figure~\ref{fig:M31_DM_View} are two notable, highly extended gas clouds in the direction of M31, namely, Complex H~\cite{hulsbosch1975studies,blitz1999high,Lockman:2003zs,Simon:2005vh} and the M31 cloud~\cite{blitz1999high,kerp2016survey}. The gas contours show \hi\ emission from the HI4PI all-sky survey (based on EBHIS and GASS)~\cite{bekhti2016hi4pi}. The M31 cloud is a highly extended lopsided gas cloud centered in projection on M31, originally reported in Ref.~\cite{blitz1999high}. It remains uncertain whether the M31 cloud resides in M31 or the MW. Most recently Ref.~\cite{kerp2016survey} has argued that M31's disk is physically connected to the M31 cloud. If at the distance of M31 ($\sim$785 kpc) the total gas mass is estimated to be $\sim10^{8}$--$10^{9} \ \mathrm{M_\odot}$. Complex H can be seen toward the top of M31's DM halo. The distance of Complex H from the MW is uncertain, although its likely distance has been estimated to be $\sim$30 kpc from the GC, which corresponds to the cloud having a diameter of about $\sim$10 kpc and an \hi\ mass of $\sim10^7  \ \mathrm{M_\odot}$~\cite{blitz1999high,Lockman:2003zs,Simon:2005vh}. Complex H does not appear to contain any stars, and it has been postulated to be either a dark galaxy of the Local Group or an example of a cold accretion flow~\cite{Simon:2005vh}. 

Figure~\ref{fig:M31_DM_View} also shows \hi\ emission contours corresponding to M33. $\gamma$-ray emission from M33 has recently been detected~\cite{karwin2019fermi,2020ApJ...894...88A,2020arXiv200307830X}, making it the only extragalactic satellite galaxy to be detected in $\gamma$ rays. The total \hi\ mass of the M33 disk is $\sim10^9  \ \mathrm{M_\odot}$. The hook-shaped gas cloud to the right of M33 is Wright's cloud, first reported in Ref.~\cite{wright1979tail}. The distance of Wright's cloud remains uncertain~\cite{Braun:2003ey}. The \hi\ mass of Wright's cloud at the distance of M33 is $\sim4.5 \times 10^7  \ \mathrm{M_\odot}$~\cite{keenan2015arecibo}. Although no contours are shown, we note that below M33 is ``the dark companion to M33", which is another highly extended gas cloud originally reported in Ref.~\cite{thilker2002high}, and labeled as a compact high-velocity cloud. If at the distance of M33, Ref.~\cite{keenan2015arecibo} estimates the \hi\ mass to be $\sim10^7  \ \mathrm{M_\odot}$, and the size to be $\sim18.2 \times 14.6 \ \mathrm{kpc}$. See~\cite{keenan2015arecibo} for details of the cloud. 

The main objective of Figure~\ref{fig:M31_DM_View} is to provide a qualitative summary of some well-known objects in the line of sight towards M31's outer halo. In particular, for the M31 satellites we do not necessarily expect to detect them individually in $\gamma$ rays (aside from M33). For the gas clouds, any $\gamma$-ray emission would depend on the actual location of the cloud, along with the CR density in the region. To investigate this in depth would require a detailed modeling which is beyond the scope of this analysis.

\section{Analysis}
\label{sec:analysis}
\subsection{Gamma-Ray Data for M31}

\begin{figure*}
\centering
\includegraphics[width=0.45\textwidth]{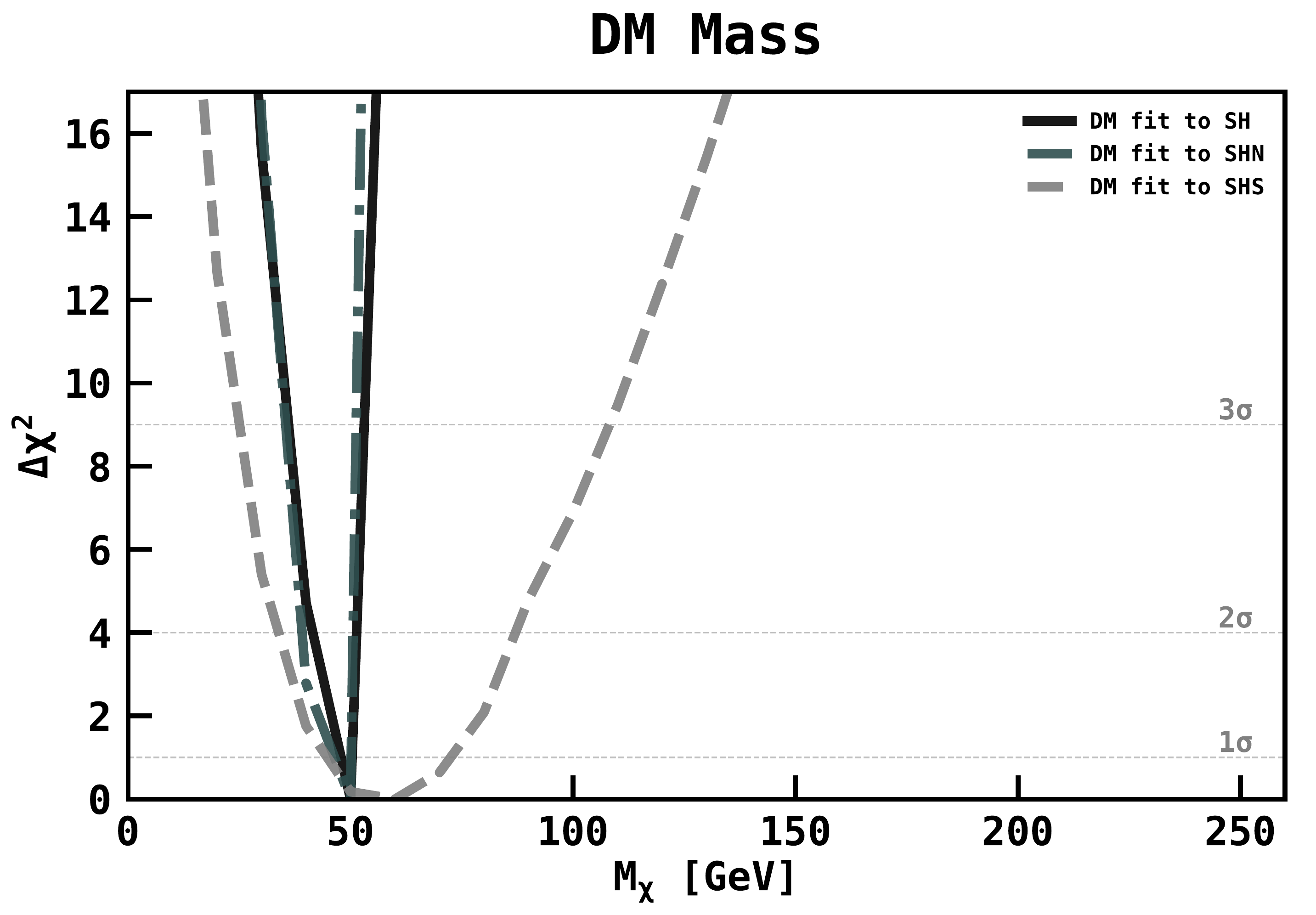}
\includegraphics[width=0.48\textwidth]{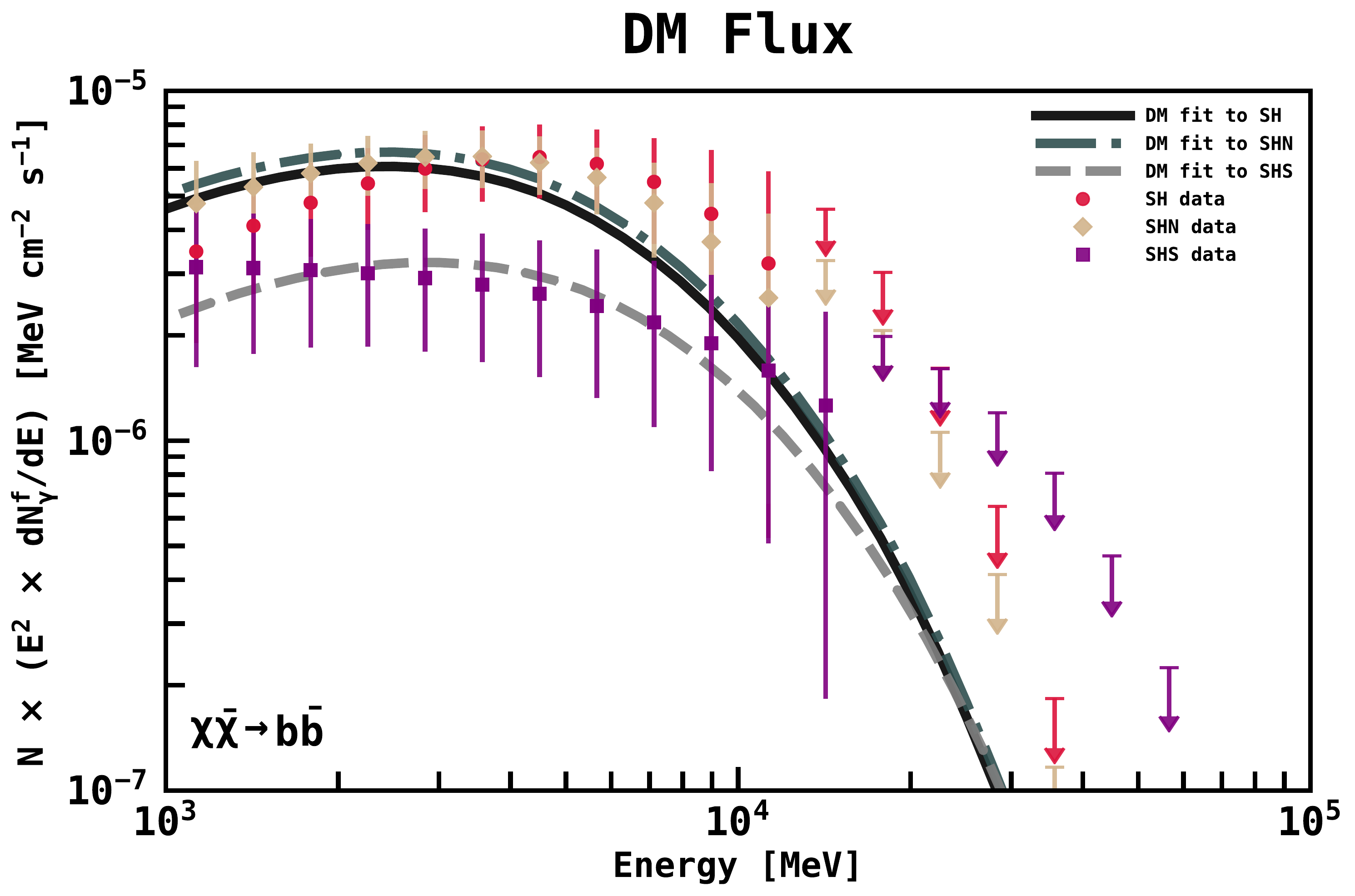}
\caption{\textbf{Left panel}: $\mathrm{\Delta \chi^2}$ profile for the three different fit variations: spherical halo (SH): solid black curve; spherical halo north (SHN): dash-dot turquoise curve; spherical halo south (SHS): dashed grey curve. The light grey dotted lines show the 1, 2, and 3 sigma contour levels, for 1 degree of freedom. \textbf{Right panel}: Best-fit spectra overlaid to the corresponding data. Arrows give the 1$\sigma$ upper limits.}
\label{fig:M31_radial_profile}
\end{figure*}

To determine whether the excess $\gamma$-ray emission observed towards M31's outer halo is consistent with a DM interpretation, we employ the best-fit $\gamma$-ray spectra from Ref.~\cite{karwin2019fermi}. In that study M31's halo is characterized using three symmetric components centered at M31 labeled as: inner galaxy (IG; r $\leq 0.4^\circ$), spherical halo (SH; $0.4^\circ >$ r $\leq 8.5^\circ$), and far outer halo (FOH; r $> 8.5^\circ$). For an M31-MW distance of 785 kpc, the IG, SH, and FOH correspond to projected radii of 5.5 kpc, 117 kpc, and $\sim$200 kpc, respectively. In this paper we only consider the SH component. The IG component is complicated by uncertainty in the expected $\gamma$-ray emission from standard astrophysical processes. The FOH component overlaps with the MW plane at the top of the field, which significantly complicates the interpretation of the emission from this region. In addition, properly modeling the FOH will require a thorough treatment of secondary emission from DM, which we leave for a future study.

Two different fit variations were performed in Ref.~\cite{karwin2019fermi} to determine the spectrum of the SH component. In the main variation (full) the entire template was used. In an alternative variation (north and south) the template was separated into north and south components. In this case the spectral parameters for the two halves are allowed to vary independently, although they are fit simultaneously. This results in three different determinations of the spectrum, which we label as spherical halo (SH), spherical halo north (SHN), and spherical halo south (SHS). We use these variations to quantify the systematic uncertainty of the signal related to modeling the MW foreground, which differs in the two regions. 

It is important to emphasize that the line of sight towards M31 extends through the MW DM halo, in addition to the M31 DM halo. However, the potential $\gamma$-ray contribution from the MW component is not explicitly accounted for when determining the M31 contribution. Some of the MW halo component would likely be attributed to the isotropic component, as well as to the other components of the IEM; however, it is unclear the extent to which this would occur. This is partly due to the fact that the absorption of a MW DM halo signal by other MW components in large part depends on the actual halo geometry and substructure content in the direction of the M31 field. Thus the spectra for the M31-related components from Ref.~\cite{karwin2019fermi} contain the total excess emission along the line of sight, which may also include some significant contribution from the MW's extended DM halo. This is taken into account in our $J$-factor calculations.

\subsection{Dark Matter Fit}

As our representative DM model we consider annihilation into bottom quarks. This channel has been shown to provide a good fit to the $\gamma$-ray GC excess. The DM spectra\footnote{available at \url{http://www.marcocirelli.net/PPPC4DMID.html}} are obtained from PPCC 4 DM ID~\cite{Cirelli:2010xx,Ciafaloni:2010ti}, and they include electroweak corrections. We scan DM masses from 10 GeV to 280 GeV, using a 10 GeV spacing. 

The $\gamma$-ray flux for DM annihilation is given by

\begin{equation}
\label{eq1}
\frac{d\Phi}{dE} =  \frac{<\sigma_f v >}{4 \pi \eta m_\chi^2} \frac{dN_\gamma^{f}}{dE} J,
\end{equation}

where $<\sigma_f v >$ is the velocity averaged annihilation cross-section for final state $f$, $m_\chi$ is the DM mass, $\eta=$ 2 (4) for self-conjugate (non-self-conjugate) DM, $dN_\gamma^{f}/dE$ is the number of $\gamma$-ray photons for annihilation into final state $f$, and $J$ is the astrophysical $J$-factor, which will be discussed in Section~\ref{sec:J-factor}. In general Eq.~\eqref{eq1} is summed over all final states $f$. In this analysis we use $\eta=$ 2. 

By multiply each side of Eq.~\eqref{eq1} by the energy squared we obtain units of $\mathrm{MeV \ cm^{-2} \ s^{-1}}$:
$$
E^2\frac{d\Phi}{dE} =  \frac{<\sigma_f v >}{4 \pi \eta m_\chi^2} (E^2\frac{dN_\gamma^{f}}{dE}) J.
$$

To fit to the $\gamma$-ray data we freely scale the quantity in parentheses by a normalization factor $N$, using a $\chi^2$ fit. This then implies:

\begin{equation}
\label{eq2}
N =  \frac{<\sigma_f v > }{4 \pi \eta m_\chi^2}J.
\end{equation}
 
The M31 data contains upper limits which need to be accounted for in the fit procedure. For $n$ measurements of $x_i$ with uncertainties $\sigma_i$ and $m$ upper limits with $x_j < n \sigma_j$ ($n$th confidence level), the $\chi^2$ can be defined as~\cite{2016ApJ...816...85L,1986ApJ...306..490I}

\begin{equation}
\label{eq7}
\chi^2 = \sum_i^n z_i^2 - \sum_j^m 2 \mathrm{ln} \frac{1 + \mathrm{erf} (z_j / \sqrt{2})}{2},
\end{equation}
where
\begin{equation}
\label{eq8}
 z_i = \frac{x_i - \hat{x}_i(\theta)}{\sigma_i},
\end{equation}
and
\begin{equation}
\mathrm{erf}(x) = \frac{2}{\sqrt{\pi}}\int_0^x e^{-t^2}dt.
\end{equation}
 
The first term on the right-hand side in Eq.~\eqref{eq7} is the classic definition of chi-squared, and the second term introduces the error function to quantify the fitting of upper limits. The quantity $\hat{x_i}(\theta)$ in Eq.~\eqref{eq8} is the modeled value. We also calculate the reduced chi-squared:
\begin{equation}
\chi^2_{\mathrm{red}} = \frac{\chi^2}{\nu},
\end{equation}
with the degrees of freedom  $\nu = 20 - 1 = 19$, corresponding to 20 energy bins and 1 free parameter in the fit. 
 
Results for the fit are shown in Figure~\ref{fig:M31_radial_profile}. The left panel shows the $\mathrm{\Delta \chi^2}$ profile for the three different fit variations. Dashed grey lines show the 1,2, and 3 sigma contour levels (for 1 degree of freedom), corresponding to $\mathrm{\Delta \chi^2}$ values of 1, 4, and 9, respectively. The best-fit mass for the SH model is $50^{+0.4}_{-2.3}$ GeV, with $\mathrm{\chi^2_{red}} = $ 1.03, and $N = (6.2 \pm 0.6) \times 10^{-10}$. The best-fit mass for the SHN model is $50^{+0.2}_{-4.0}$ GeV, with $\mathrm{\chi^2_{red}} = $ 0.9, and $N = (6.8 \pm 0.5) \times 10^{-10}$. And the best-fit mass for the SHS model is $60^{+13.2}_{-15.9}$ GeV, with $\mathrm{\chi^2_{red}} = $ 0.5, and $N = (2.8 \pm 0.4) \times 10^{-10}$. The corresponding best-fit spectra are plotted in the right panel of Figure~\ref{fig:M31_radial_profile}, overlaid to the corresponding data. 

\subsection{Analytical Determination of the J-Factor}
\label{sec:J-factor}
For the best-fit models the corresponding annihilation cross-section is calculated using Eq.~\eqref{eq2}. This requires knowledge of the $J$-factor, which is the greatest uncertainty in the analysis. The $J$-factor characterizes the spatial distribution of the DM, and is given by the integral of the mass density squared, over the line of sight. When describing the DM distribution as an ensemble of disjoint DM halos, the $J$-factor is:
\begin{equation} \label{eq3}
J=\sum_i\int_{\Delta\Omega}d\Omega\int_{\text{LoS}}ds\rho_i^2(\mathbf{r}_i(s,\mathbf{n})),
\end{equation}
summed over all halos in the line of sight (LoS), where $\rho_i(\mathbf{r})$ is the density distribution of halo $i$, and $\mathbf{r}_i(s,\mathbf{n})$ is the position within that halo at LoS direction $\mathbf{n}$ and LoS distance $s$. 

$J$-factors determined from these spherically-averaged profiles are an underestimate of the total $J$-factor because of the effect of the non-spherical structure. This underestimate is typically encoded with a boost factor ($B$). To calculate $J$-factors we use the CLUMPY\footnote{available at \url{https://clumpy.gitlab.io/CLUMPY/}} code~\citep{charbonnier2012clumpy,bonnivard2016clumpy,hutten2019clumpy}. For a detailed discussion of the boost factor calculation see the CLUMPY papers/website, as well as Refs.~\cite{Bullock:1999he,Springel:2008cc,Gao:2011rf,Ishiyama:2014uoa,Moline:2016pbm} and references therein. Here we summarize the key points. The main parameters for the boost factor are the following:

\begin{itemize}
\item[$\circ$] minimum subhalo mass
\item[$\circ$] mass-concentration relationship
\item[$\circ$] subhalo mass function (index and normalization), i.e. the number of subhalos per volume in a given mass range 
\item[$\circ$] mass distribution of subhalos
\item[$\circ$] distribution of subhalos in the main halo
\end{itemize}

Since the $\gamma$-ray flux from DM annihilation scales as the square of the DM density, the effect of substructure is very important for indirect detection, as it provides a boost to the total flux. The flux enhancement is most significant for larger halos, since they enclose more levels of hierarchical formation. The size of the smallest DM subhalo is determined by the free streaming scale of the DM particles~\cite{Green:2003un,Ishiyama:2014uoa,ishiyama2019abundance}. This depends on the specific particle physics and cosmological models, and in general it is highly uncertain. In this study we consider minimum subhalo masses in the range $M_{\mathrm{min}} = 10^{-6} - 10^{6} \ M_\odot$. The lower limit is typically expected for thermal WIMP DM with a mass of $\sim$100 GeV~\citep{Green:2003un}, and the upper limit reflects the typical resolution power of DM simulations. 

\begin{figure}[th!]
\centering
\includegraphics[width=0.45\textwidth]{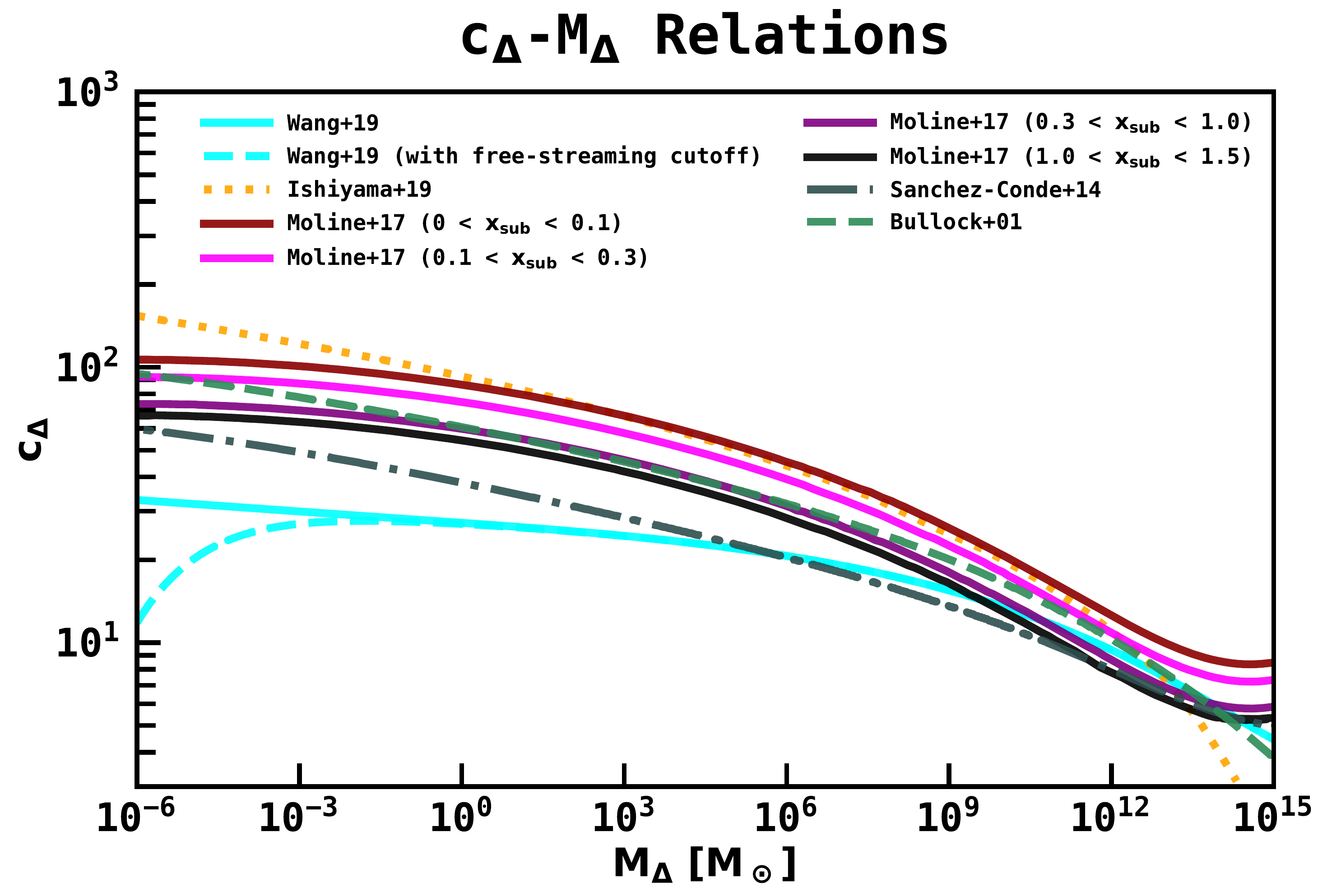}
\includegraphics[width=0.43\textwidth]{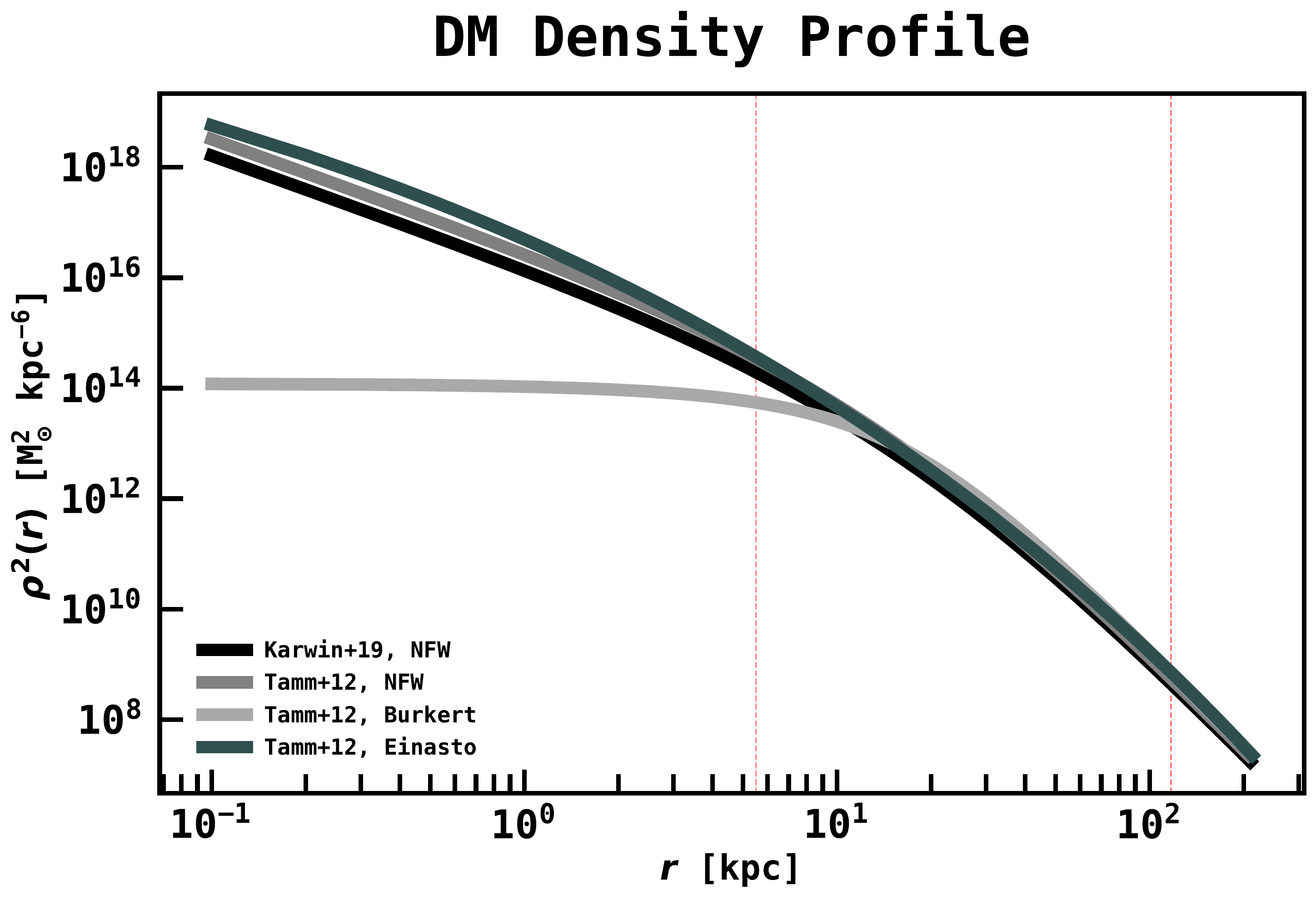}
\includegraphics[width=0.45\textwidth]{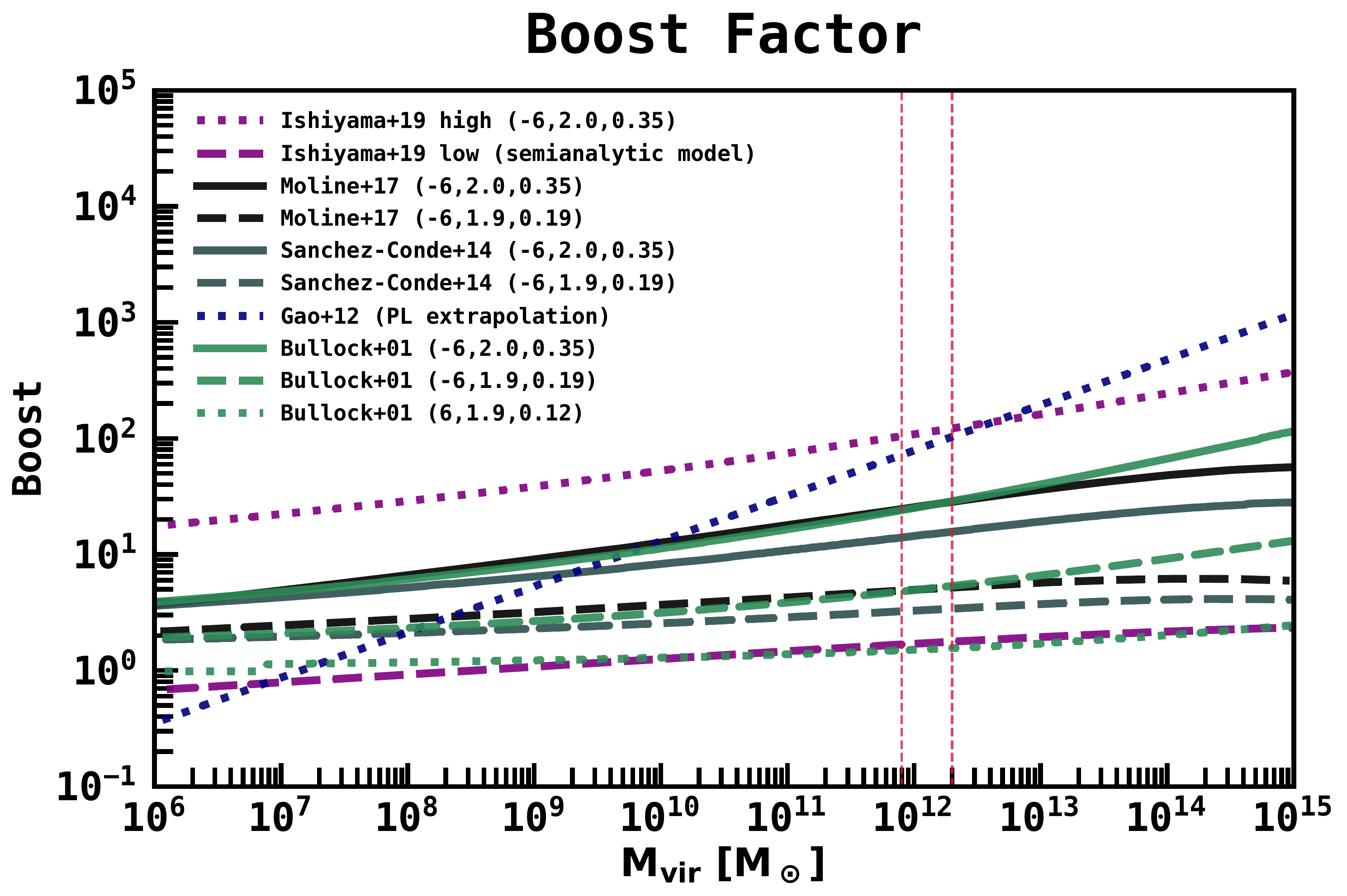}
\caption{\textbf{Top panel:} Concentration-mass relations from Refs.~\cite[solid black, purple, magenta, and red]{Moline:2016pbm},~\cite[dotted yellow]{ishiyama2019abundance},~\cite[dash-dot grey]{sanchez2014flattening},~\cite[dashed green]{Bullock:1999he}, and~\cite[solid and dashed cyan]{2019arXiv191109720W}. \textbf{Middle panel:} Different DM density profiles for M31. The region bounded by the red dashed lines corresponds to the SH. \textbf{Bottom panel:} Mass dependence of the boost factor for different parameters. The name in the legend specifies the model of the concentration-mass relation, and in parentheses the numbers give (in order) the power of the minimum subhalo mass, the PL index of the subhalo mass function, and the fraction of the halo resolved in substructure. The red dashed lines correspond to the mass range for M31 and the MW.}
\label{fig:Boost_factor_values}
\end{figure}

The concentration parameter $c_\Delta$, at a given characteristic overdensity $\Delta$, can be defined as
\begin{equation}
c_\Delta = \frac{R_\Delta}{r_{-2}},
\end{equation}
where $R_\Delta$ is the radius of the DM halo corresponding to the overdensity $\Delta$, and $r_{-2}$ is the position where the slope of the DM density profile reaches $-$2. The boost factor is highly sensitive to the concentration parameter, as it scales as the concentration to the third power~\citep{charbonnier2012clumpy,bonnivard2016clumpy,hutten2019clumpy}. In general the concentration is a function of halo mass and redshift. In the top panel of Figure~\ref{fig:Boost_factor_values} we plot different determinations of the concentration-mass relation at z=0. The solid lines (black, purple, magenta, and red) are from Ref.~\cite{Moline:2016pbm}, which is based on two N-body cosmological simulations of MW-sized haloes: VL-II~\cite{diemand2008clumps} and ELVIS~\cite{garrison2014elvis}. These results summarize some of the main properties of the concentration parameter; namely, for a given halo the concentration decreases with increasing radius, and the concentration of subhalos is higher than that of field halos. In particular, the solid lines in Figure~\ref{fig:Boost_factor_values} are for different radial bins defined in terms of $x_{\mathrm{sub}}\equiv R_\mathrm{sub}/R_\Delta$. The solid black line is calculated outside of the virial radius, and it gives an approximation for field halos (see~\cite{Moline:2016pbm} for further details). For simplicity, in our benchmark model we use the relation from Ref.~\cite{Bullock:1999he}, plotted with a dashed green line in the top panel of Figure~\ref{fig:Boost_factor_values}. As can be seen, this serves as a good intermediate model between the different estimates. Note that we have also tested the model from Ref.~\cite{sanchez2014flattening} and the results are qualitatively consistent. 

The boost factor also depends on the subhalo mass function, which specifies the number of subhalos at a given mass. This function is given by a simple power law (PL), having an index of $\sim-1.9$ to $-2.0$~\citep{Springel:2008cc,Moline:2016pbm}. The normalization of the PL is chosen so that the mass of the DM halo resolved in substructure is a specified amount. To bracket the uncertainty in the $J$-factor for both M31 and the MW, we vary the index of the subhalo mass function ($\alpha$) and the fraction of the halo resolved in substructure ($f_{\mathrm{sub}}$) in the ranges $1.9-2.0$ and $0.12-0.35$, respectively. These values are representative of the current uncertainty~\cite{Springel:2008cc,Diemand:2006ik,DiMauro:2019frs}.  

\begin{figure*}[t]
\centering
\includegraphics[width=0.32\textwidth]{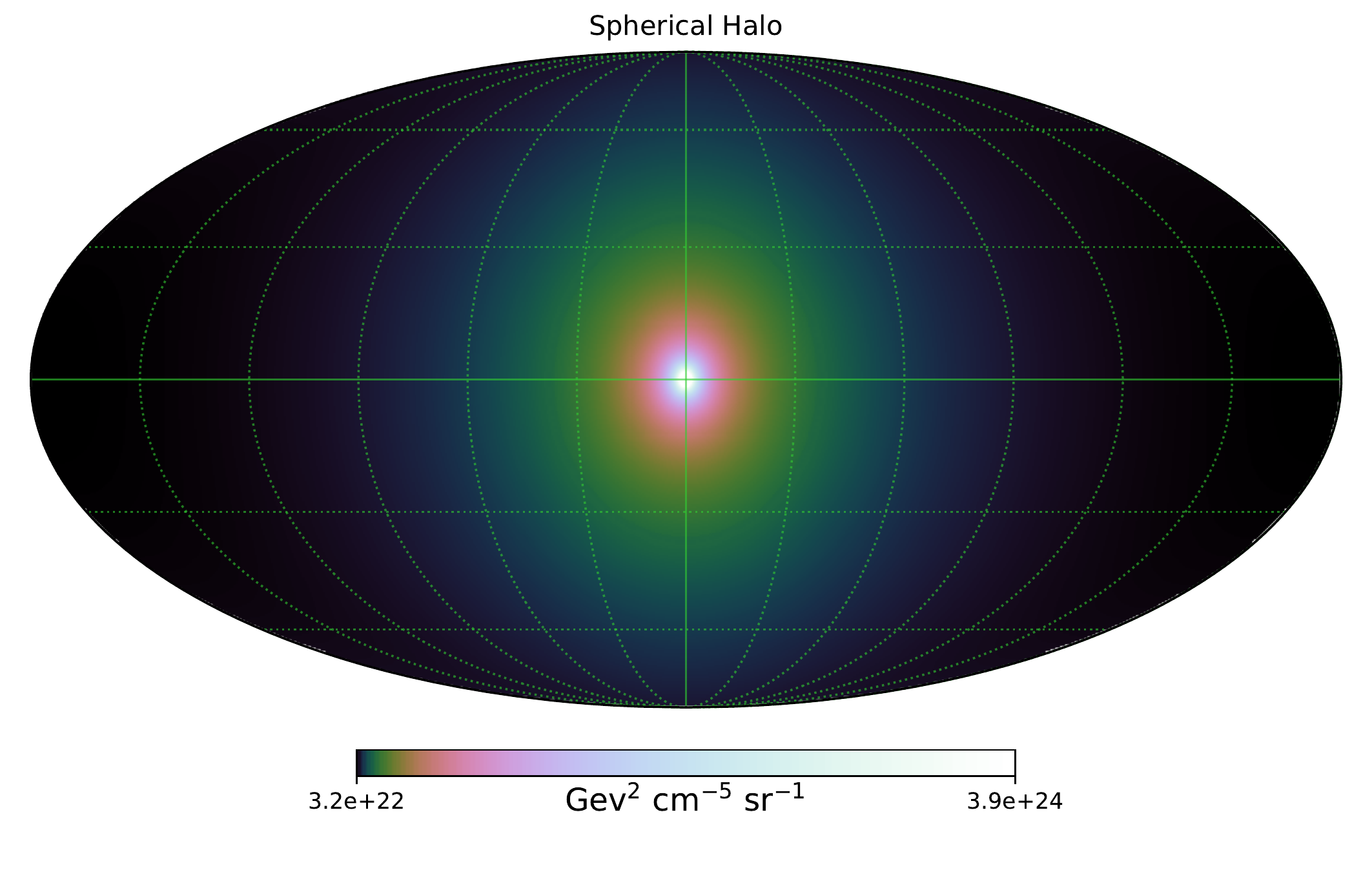}
\includegraphics[width=0.32\textwidth]{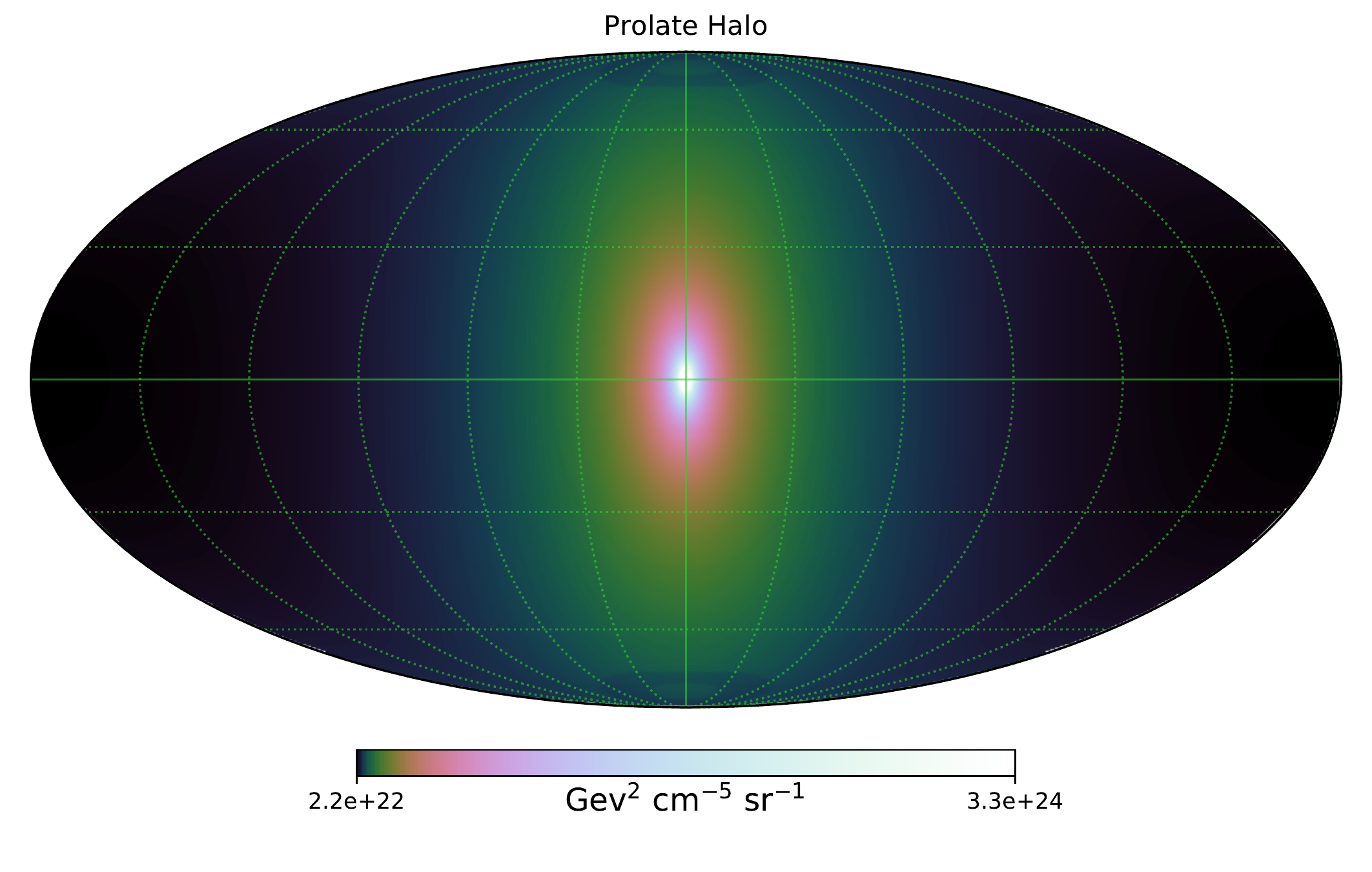}
\includegraphics[width=0.32\textwidth]{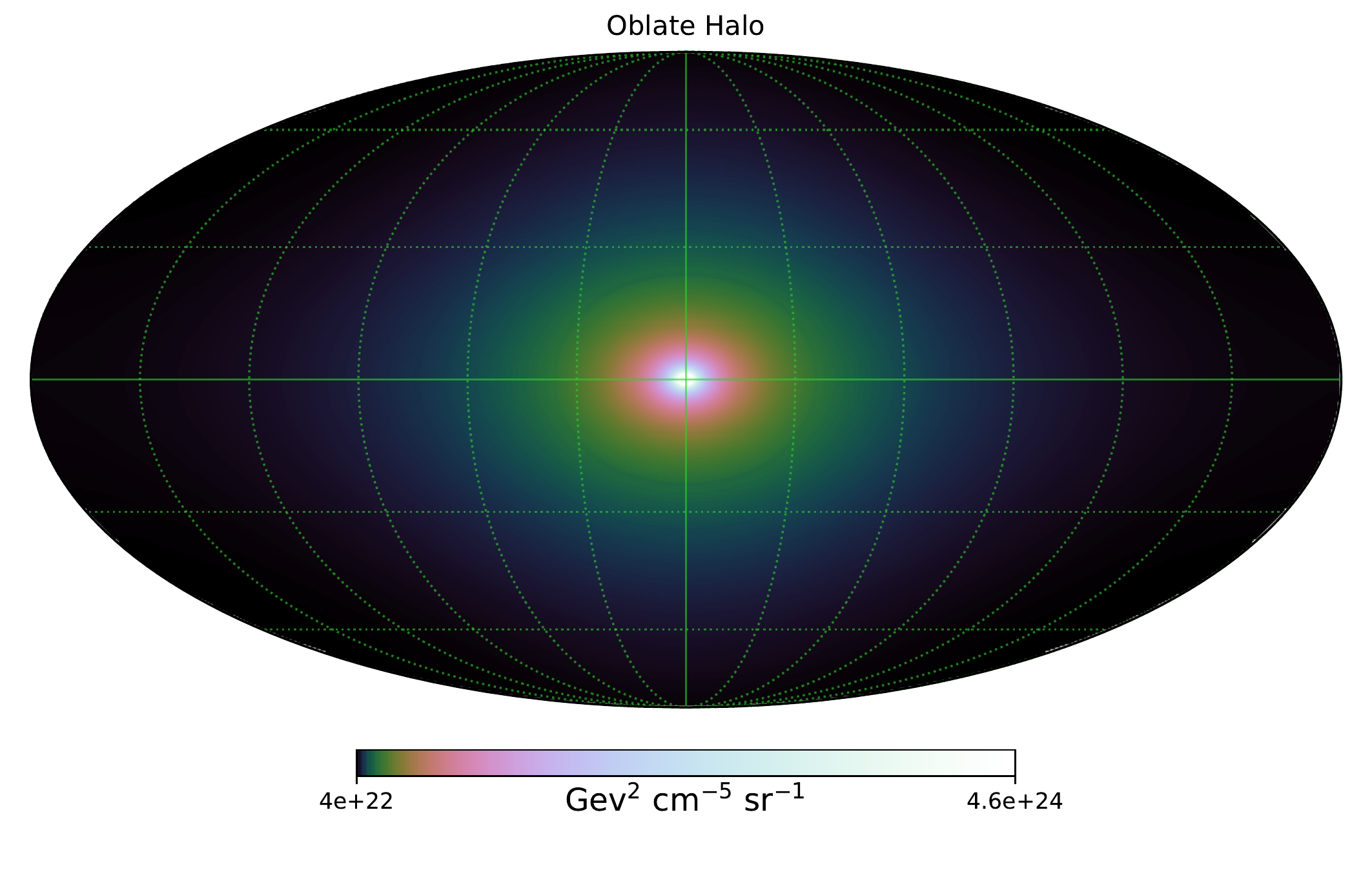}
\caption{MW $J$-factors for four different geometries, as indicated above each map. Maps are shown in Galactic coordinates with a Mollweide projection. The corresponding axis ratios are given in Table~\ref{tab:geometry}. For the prolate halo q=1.67, and for the oblate halo q=0.6. The color scale ranges from the minimum halo value to 1/10 the maximum halo value. The DM model is "Einasto high" from Table~\ref{tab:J_and_Cross}. Note that these particular maps don't show individually resolved substructures, although they are included in the analytical model.}
\label{fig:geometry} 
\end{figure*}

The middle panel of Figure~\ref{fig:Boost_factor_values} shows different DM density profiles for M31. The region bounded by the dashed red lines corresponds to the SH, where the fit to the $\gamma$-ray data is performed. The solid black curve is from Ref.~\cite{karwin2019fermi}, and the other curves are from Ref.~\cite{tamm2012stellar}. For our $J$-factor calculations we test two profiles. We use the NFW profile from Ref.~\cite{karwin2019fermi}, which has corresponding halo properties of $R_{\mathrm{vir}} = 210$ kpc, $R_\mathrm{s} = 18.9$ kpc, and $\rho_\mathrm{s} = 2 \times 10^6 \ M_\odot \ \mathrm{kpc}^{-3}$. In CLUMPY this corresponds to the kZHAO profile with parameters $\alpha,\beta,\gamma$ = 1,3,1. We also use the Einasto profile from Ref.~\cite{tamm2012stellar}, which has the corresponding halo properties of $R_{\mathrm{vir}} = 210$ kpc, $R_\mathrm{s} = 178$ kpc, and $\rho_\mathrm{s} = 8.12 \times 10^3 \ M_\odot \ \mathrm{kpc}^{-3}$. In CLUMPY this corresponds to the kEINASTO\_N profile with the parameter n=6. The overdensity factor is set to $\Delta=200$. We use an M31-MW distance of 785 kpc. 

Other major uncertainties in the boost factor calculation are the spatial distribution of subhaloes in the main halo, as well as the mass distribution of the subhaloes themselves. We assume that the density profile and the spatial distribution of the subhaloes are the same as the density profile of the main halo for both the NFW and Einasto distributions. Note that both the spatial distribution of subhaloes and their density profiles have been found to prefer an Einasto distribution compared to an NFW, although both profiles provide a good fit (see \cite{Springel:2008cc} and references therein). Additionally, it's found that within $\sim$25 kpc from the center of MW-sized halos there is a depletion of the subhalo population due to tidal disruption from the galactic disk~\cite{garrison2017not}.

In principle each DM halo of a given mass is a hierarchical structure, so that even subhalos have subhalos themselves. For simplicity we set the number of substructure levels to 2. We have also tested including higher substructure levels, but we find that they do not make a significant difference for our $J$-factor calculations, as has been previously found~\cite{Moline:2016pbm}. 

The bottom panel of Figure~\ref{fig:Boost_factor_values} shows the mass dependence of the boost factor for different choices of the minimum subhalo mass, the subhalo mass function, and the fraction of the halo mass resolved in substructure. Within the uncertainties we have considered, the overall boost factor ranges from $\sim$1.5--26.0 (for an NFW density profile). Note that this is the value reported by CLUMPY for the entire halo, which we report here for easy comparison with different values from the literature. 

\subsection{Halo Geometry}
\label{sec:halo_geometry}

Another important systematic uncertainty for determining the $J$-factor for the M31 field is the halo geometry, for both M31 and the MW. Indirect DM searches typically assume spherical symmetry for the halo shape, however, in the standard DM paradigm ($\Lambda$CDM), DM halos are expected to be very non-spherical, and in fact, spherical halos are rare (see~\cite{Allgood:2005eu} and references therein). 

For the MW, numerous studies have been done to infer the DM halo geometry, but differing conclusions have been reached. The halo has been found to be spherical~\cite{2019MNRAS.485.3296W}, prolate~\cite{Bowden_2016,Banerjee:2011rr,2019A&A...621A..56P}, oblate~\cite{2014ApJ...794..151L}, triaxial (including the so-called ``Gaia sausage")~\cite{Law:2009yq,2019MNRAS.482.3868I,Evans:2018bqy}, and even lopsided~\cite{Saha:2009dt}. Further complicating the matter is that the halo geometry may have a radial dependence~\cite{2020arXiv200909220E,2020arXiv200910726V}. Moreover, it's found in both simulations and observations that for galaxy pairs (similar to M31 and the MW) the halos tend to bulge toward their respective partners~\cite{Conn:2013iu,pawlowski2017lopsidedness}.

In general the halo geometry can be described with an ellipsoid, with the axes a, b, and c. The shape is characterized by the axis ratios, with the normalization condition abc = 1 (see the CLUMPY code for more details). For describing the MW halo, the a-axis corresponds to the Galactic x-axis (connecting the Sun to the Galactic center), the b-axis corresponds to the Galactic y-axis, and the c-axis corresponds to the Galactic z-axis (perpendicular to the Galactic plane). We use the references cited above to calculate $J$-factors for different MW halo geometries. Note that we also consider a triaxial halo geometry modeled after the Gaia sausage. Although the evidence indicates that this structure may be a subdominant component of the halo, for simplicity we test a more extreme scenario where the entire halo follows this geometry. Figure~\ref{fig:geometry} shows the three main halo shapes that we test, and the specific axis ratios for all geometries are summarized in Table~\ref{tab:geometry}. 

\begin{figure}
\centering
\includegraphics[width=0.40\textwidth]{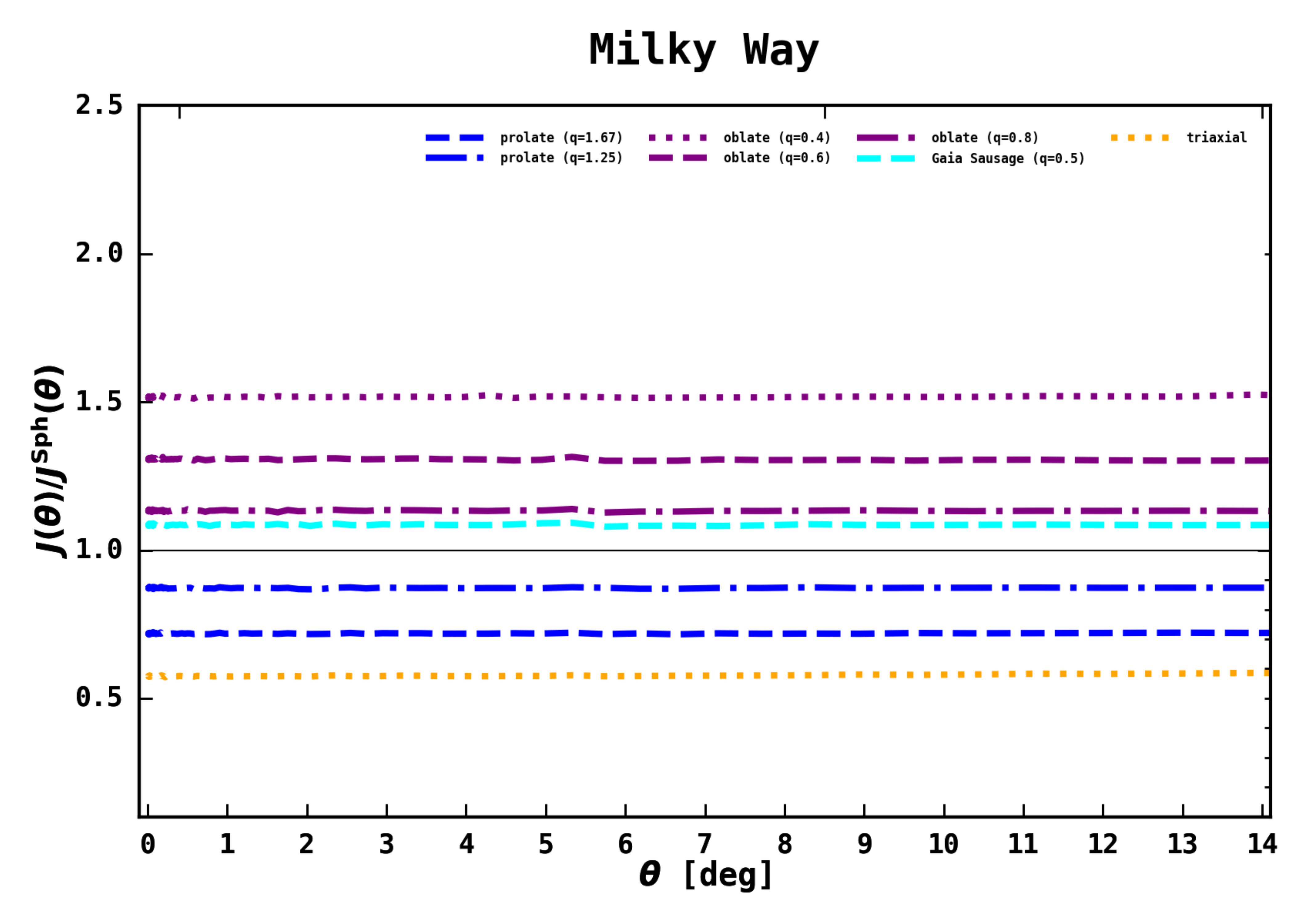}
\includegraphics[width=0.40\textwidth]{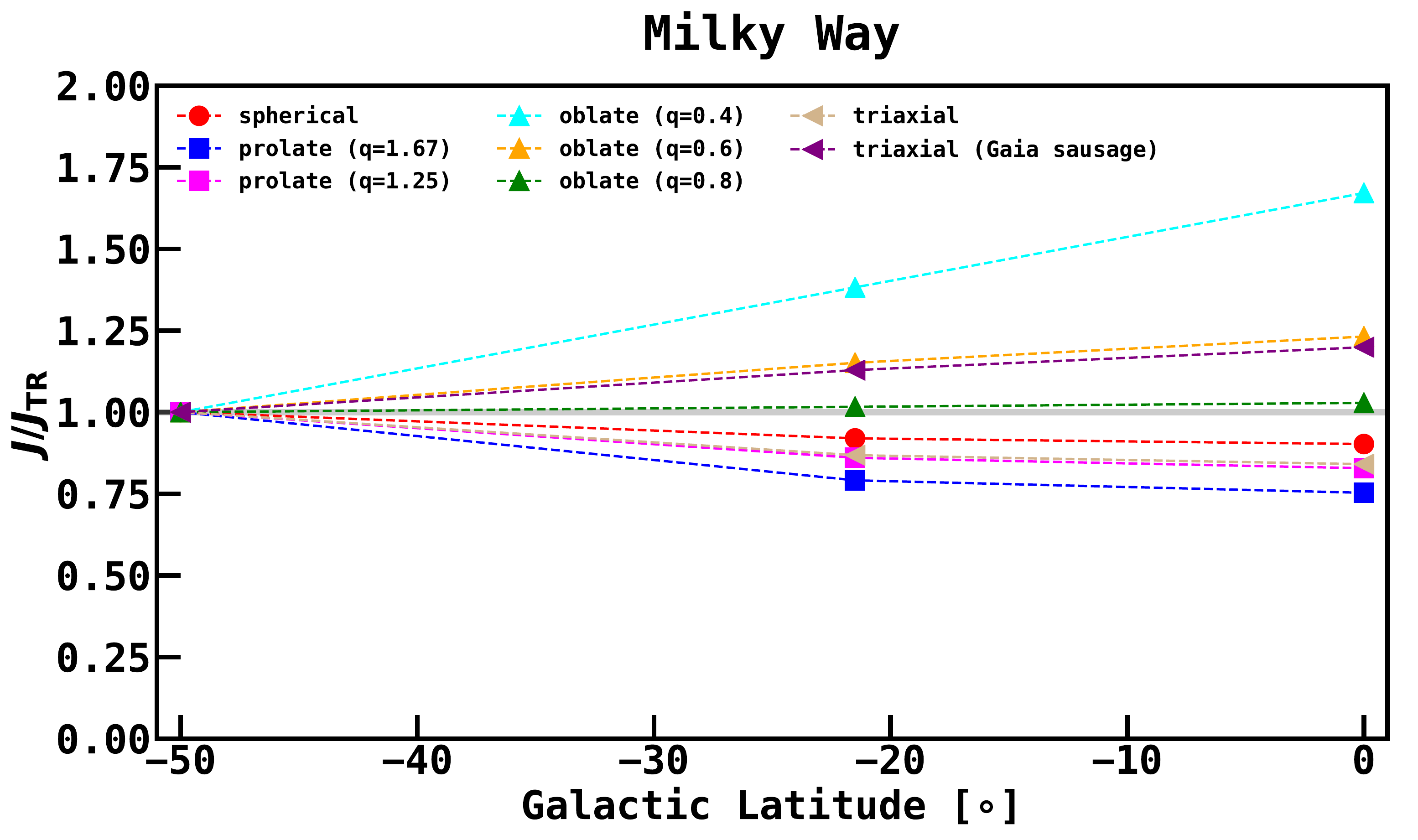}

\includegraphics[width=0.40\textwidth]{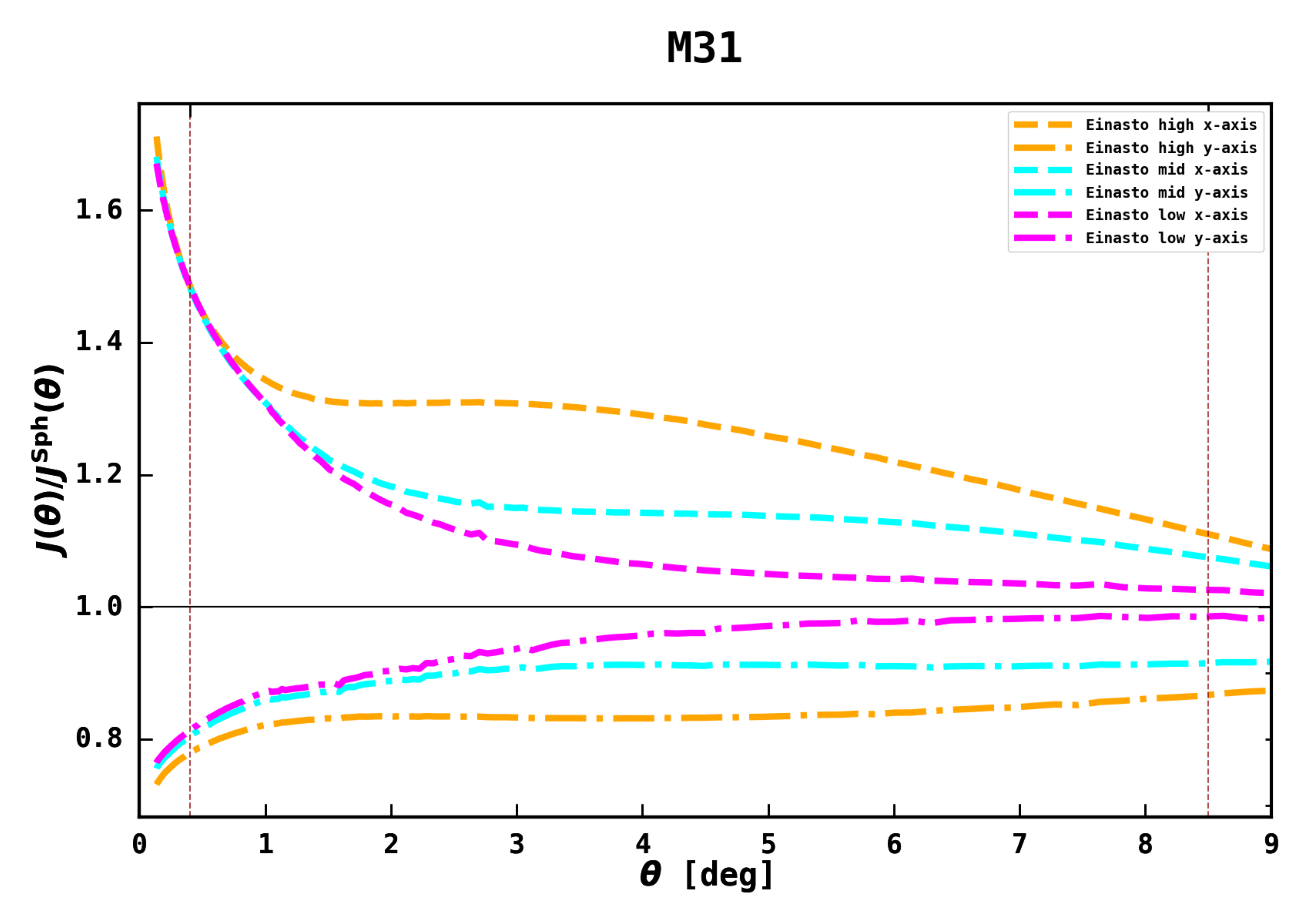}
\caption{\textbf{Top:} Ratio of the $J$-factor ($J$) for different MW halo geometries compared to a spherical halo ($J^{\mathrm{Sph}}$), for an Einasto density profile. \textbf{Middle:} Gradient ratio for the $J$-factor calculated with the line of sight centered at three different Galactic latitudes (with $l=121^\circ$). The ratio is calculated with respect to a latitude of $b=-50^\circ$ ($J_{\mathrm{TR}}$), which is comparable to the region used for tuning the isotropic spectrum in Ref.~\cite{karwin2019fermi}. The middle data points at $l=--21.5^\circ$ correspond to the M31 field. In all cases the $J$-factors are integrated over the region $0.4^\circ$ to $8.5^\circ$, using the Einasto high model from Table~\ref{tab:J_and_Cross}. \textbf{Bottom:} Ratio of the $J$-factor ($J$) for different M31 halo geometries compared to a spherical halo ($J^{\mathrm{Sph}}$), for an Einasto density profile.}
\label{fig:gradient} 
\end{figure}

 In the top panel of Figure~\ref{fig:gradient} we show the $J$-factor ratio ($J/J_{\mathrm{Sph}}$) for the Einasto high DM model, where $J$ is for the alternative geometry, and $J_{\mathrm{Sph}}$ is for the spherical halo. The ratio range for all DM models is given in Table~\ref{tab:J_and_Cross}. We find that at most the halo shape may increase or decrease the MW $J$-factor (with respect to spherical geometry) by factors of 2.29 and 0.34, respectively. 

\begin{table}
\begin{ruledtabular}
\caption{MW Halo Geometry}\label{tab:geometry}
\begin{tabular}{lcc}
Halo Geometry&Axes (a,b,c)   \\ 
\midrule 
 Spherical &1, 1, 1 \\
 Prolate (q=1.67) &0.84, 0.84, 1.41 \\
 Prolate (q=1.25)&0.93, 0.93, 1.16 \\
 Oblate (q=0.4)&1.36, 1.36, 0.54 \\
 Oblate (q=0.6)&1.19, 1.19, 0.71 \\
 Oblate (q=0.8)&1.08, 1.08, 0.86 \\
 Triaxial &0.67, 1.34, 1.113 \\
 Triaxial (Gaia Sausage, $\alpha=70^{\circ}$)&1.38, 1.06, 0.69 \\
 \end{tabular}
\end{ruledtabular}
\begin{tablenotes}

\item Note: The axes are normalized so that abc=1. In general, prolate halos have a=b$<$c, and oblate halos have a=b$>$c. For convenience we also give the ratio q=c/a. The specific axis ratios come from the literature, as discussed in the text. For visualization purposes, the different geometries are plotted in Figure~\ref{fig:geometry}.
\end{tablenotes}
\end{table}

\begin{table*}
\begin{ruledtabular}
\caption{J-Factors and Cross Sections}\label{tab:J_and_Cross}
\begin{tabular}{lcccccccccc}
Model& $\alpha_{\rm{sub}}$ & $f_{\rm{sub}}$  &\thead{$M_{\rm{min}}$ \\ $[\rm{M_\odot}]$} & \thead{$J_{\rm{MW}}$ \\ $(\times 10^{20})$ \\ $\rm{[GeV^2 \ cm^{-5}]}$} & \thead{$J_{\rm{M31}}$ \\ $(\times 10^{20})$  \\ $\rm{[GeV^2 \ cm^{-5}]}$} &\thead{$J/J_{\mathrm{Sph}}$ \\ (MW)}&\thead{$J/J_{\mathrm{Sph}}$ \\ (M31)}&$J_{\mathrm{MW}}/J_{\mathrm{TR}}$& \thead{ $<\sigma v>_\mathrm{I}$ \\ $(\times10^{-26})$ \\  $\rm{[cm^3 \ s^{-1}]}$}  & \thead{$<\sigma v>_{\mathrm{II}}$ \\ $(\times10^{-26})$ \\ $\rm{[cm^3 \ s^{-1}]} $}   \\ 
\midrule 
 Einasto high&2.0&0.35&$10^{-6}$&27.5&3.6&0.57, 1.52&0.82, 1.32&0.79, 1.38&1.3, 0.8&10.9, 7.0\\
 NFW high&2.0&0.35&$10^{-6}$&15.0&1.8&0.57, 1.51&0.82, 1.33&0.79, 1.38&2.3, 1.5&21.6, 13.9\\
 Einasto mid&1.9 & 0.19 & $10^{-6}$ &4.6&0.6&0.49, 1.81&0.88, 1.24&0.79, 1.34 &7.6, 4.9&68.0, 43.9\\
 NFW mid &1.9 & 0.19 & $10^{-6}$ &3.3&0.3&0.45, 1.86&0.87, 1.25&0.79, 1.34&10.6, 6.8&115.5, 74.5\\
 Eiansto low&1.9 & 0.12& $10^6$&1.94&0.1&0.35, 2.29&0.90, 1.21&0.78, 1.29&10.0, 6.4&317.4, 204.8\\
 NFW low&1.9&0.12&$10^6$&1.90&0.1&0.34, 2.19&0.89, 1.22&0.77, 1.30&19.4, 12.5&401.2, 259.0\\
 Einasto smooth &\nodata&\nodata&\nodata&1.50&0.05&\nodata&\nodata&\nodata&25.0, 16.1&725.5, 468.3\\
 NFW smooth &\nodata&\nodata&\nodata&1.6&0.05&\nodata&\nodata&\nodata&23.5, 15.2&787.0, 507.8\\

\end{tabular}
\end{ruledtabular}
\begin{tablenotes}

\item Note: $J$-factors are integrated over the spherical halo component ($0.4^\circ$ to $8.5^\circ$). The largest subhalo mass is taken to be 10\% the mass of the host halo. The calculations include 2 levels of substructure. For the M31 NFW profile $R_{\mathrm{vir}} = 210$ kpc, $R_\mathrm{s} = 18.9$ kpc, and $\rho_\mathrm{s} = 2.0 \times 10^6 \ M_\odot \ \mathrm{kpc}^{-3}$. For the M31 Einasto profile $R_{\mathrm{vir}} = 210$ kpc, $R_\mathrm{s} = 178$ kpc, and $\rho_\mathrm{s} = 8.12 \times 10^3 \ M_\odot \ \mathrm{kpc}^{-3}$. The MW profiles have the same parameters except we use the local DM density $\rho_\odot=0.4 \ \mathrm{GeV}^2 \ \mathrm{cm}^{-3}$, with a solar distance $R_\odot = 8.5$ kpc.  The overdensity factor is set to $\Delta=200$. We use an M31-MW distance of 785 kpc. The spatial distribution of subhalos and the density profile of the subhaloes is the same as the density profile of the main halo for both NFW and Einasto distributions. Columns 7 and 8 give the average uncertainty range (low, high) on the $J$-factor due to the halo geometry, with respect to a spherical halo ($J_{\mathrm{Sph}}$). Column 9 shows the $J$-factor gradient (low, high) with respect to the tuning region (TR) used in Ref.~\cite{karwin2019fermi}, which is centered at $b=-50^\circ$. There are two values for each cross-section; the first value results from fitting to the SH data (SHN data is very similar), and the second value results from fitting to the SHS data. Subscript I on the cross-section indicates case I, where $J_{\mathrm{total}} = J_{\mathrm{M31}} + J_{\mathrm{MW}}$, and subscript II on the cross-section indicates case II, where $J_{\mathrm{total}} = J_{\mathrm{M31}}$. Corresponding curves are plotted in Figure~\ref{fig:more_j_factor}. 	
\end{tablenotes}
\end{table*}

To test how the MW $J$-factor varies with Galactic latitude we repeat the calculations with the line of sight centered at latitudes of $-50^\circ$ and $0^\circ$, with longitude = 121$^\circ$. Note that $b=-50^\circ$ corresponds to the region used in Ref.~\cite{karwin2019fermi} for tuning the isotropic spectrum, which we refer to as the tuning region (TR). Results for this test are shown in the middle panel of Figure~\ref{fig:gradient} (for the Einasto high model), where we plot the $J$-factor ratio with respect to the value obtained in the TR. In all cases a gradient can be seen, with the amplitude of the variation dependent on the halo geometry. This is even true for a spherical halo, due to our position in the Galaxy at $\sim$8.5 kpc from the Galactic center. The range of gradient ratios for all DM models is given in Table~\ref{tab:J_and_Cross}. In going from high latitude to low latitude, the $J$-factors for the spherical and prolate halos decrease by a minimum factor of 0.77. Alternatively, the $J$-factors for the oblate and triaxial (Gaia sausage) halos increase by a maximum factor of 1.38. Since Ref.~\cite{karwin2019fermi} tunes the isotropic spectrum in a region below the M31 field (consistent with $l=-50^\circ$), these results show that it is not necessarily the case that the MW DM halo component would be fully absorbed by the isotropic template. Moreover, even a gradient of $\sim$20$-40 \%$ (as is found in the gradient calculation) would be a significant contribution to the total $J$-factor for the M31 field. 

We also test how the $J$-factor depends on the M31 halo geometry, with the main goal of estimating the full uncertainty range. For simplicity we test two different geometries. In each case the minor-to-major axis ratio is 0.4 (with a$>$b=c). This represents a highly flattened halo, but it has also been found for M31 in particular~\cite{Banerjee:2008kt}. We test two different orientations, one with the major axis pointing along the line of sight connecting M31 and the MW (x-axis), and the other with the major axis pointing perpendicular to the line of sight (y-axis), running from left to right in the field of view. Note that results for the z-axis orientation are similar to those of the y-axis orientation. The bottom panel of Figure~\ref{fig:gradient} shows the ratio of the $J$-factor for these different geometries compared to a spherical geometry (for the Einasto high model). The uncertainty range for all DM models is given in Table~\ref{tab:J_and_Cross}. The M31 halo geometry introduces an uncertainty in the range $0.82-1.32$, where the increase is seen for the major axis aligned with the x-axis and the decrease is seen for the major axis aligned along the perpendicular axes. 

\subsection{$J$-Factor Uncertainty from the Milky Way Foreground}
\label{sec:obs_uncertainty}

In the context of the $J$-factor uncertainty from the MW foreground, we consider two extreme cases. For case I we assume that none of the MW halo signal along the line of sight has been absorbed by the isotropic component (and other components of the IEM), and thus the total $J$-factor is the sum of the $J$-factors for the MW and M31. For case II we assume that the MW halo signal along the line of sight has been completely absorbed, and so the total $J$-factor is due only to M31. In actuality, if the observed excess is in fact related to DM then the true case is likely somewhere between the two extremes. 

\subsection{Total J-Factor Uncertainty}
\label{sec:J_and_Cross}

\begin{figure}
\centering
\includegraphics[width=0.49\textwidth]{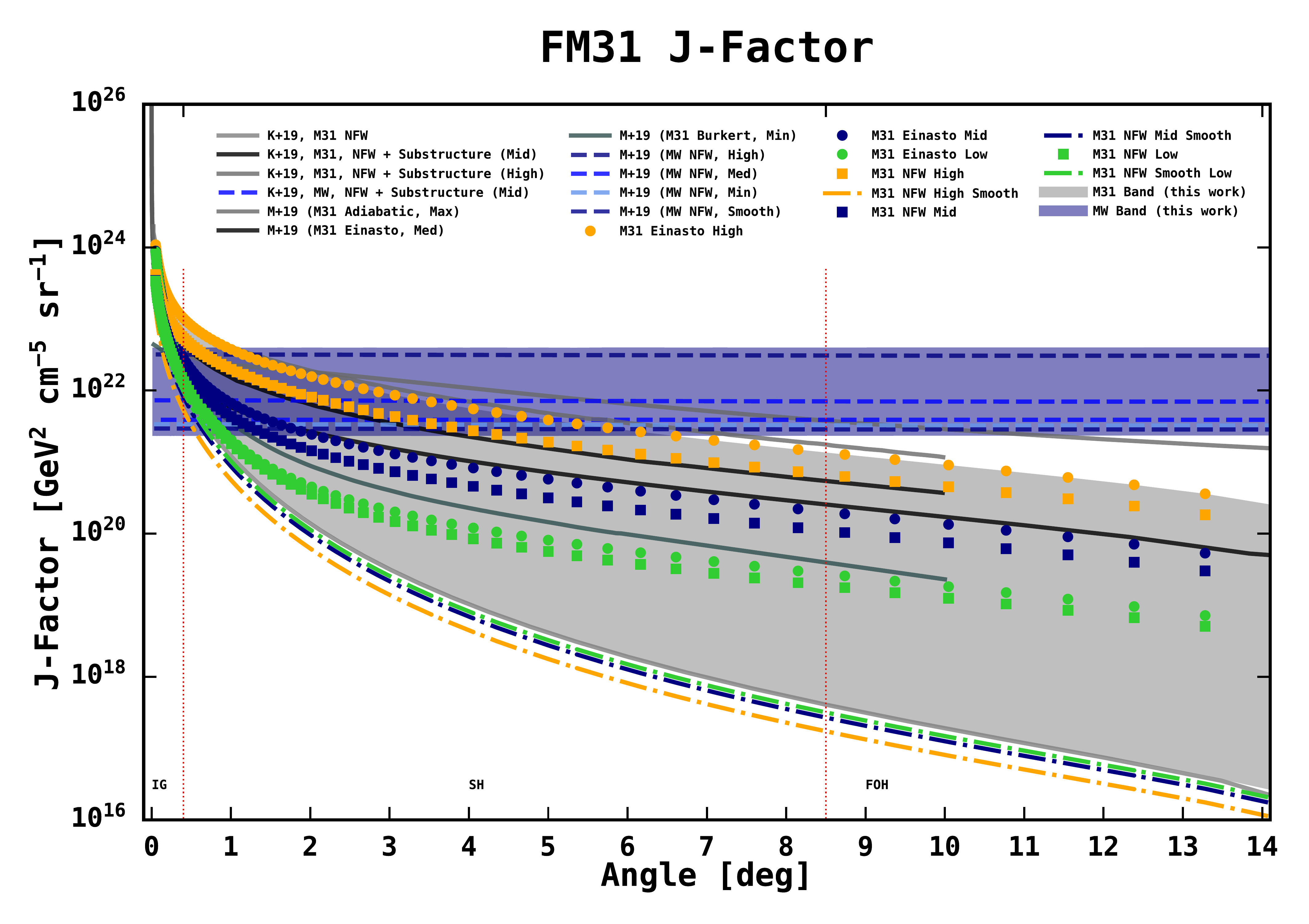}
\caption{$J$-factors for M31 and the MW. The grey band is the $J$-factor uncertainty for M31 from this work. The blue band is the $J$-factor uncertainty for the MW from this work. The markers are the M31 calculations for the NFW (squares) and Einasto (circles) profiles, with the boost factor. Parameters for the different variations are given in Table~\ref{tab:J_and_Cross}. The solid curves are independent calculations for M31 from Ref.~\cite{karwin2019fermi} (extending to 14 degrees) and Ref.~\cite{DiMauro:2019frs} (extending to 10 deg). Likewise the dashed lines are independent calculations for the MW. The dash-dot lines towards the bottom show the smooth M31 profiles corresponding to the markers. The vertical dotted red lines show the boundaries of M31's IG, SH, and FOH (the fit is performed over the SH).}
\label{fig:more_j_factor}
\end{figure}

 Figure~\ref{fig:more_j_factor} shows the different $J$-factors as a function of radial distance from the center of M31. The grey band is the $J$-factor uncertainty for M31 from this work. The purple band is the $J$-factor uncertainty for the MW from this work. The markers are the M31 calculations for the NFW (squares) and Einasto (circles) profiles, with the boost factor, corresponding to the values in Table~\ref{tab:J_and_Cross}. The dash-dot lines towards the bottom show the smooth M31 profiles corresponding to the markers. As can be seen, the smooth profiles are anti-correlated to the total profiles, i.e.~as the boost factor increases, the fraction of DM resolved in substructure also increases, and the fraction of the smooth DM component decreases. The solid curves are independent calculations for M31 from Ref.~\cite{karwin2019fermi} (extending to 14 degrees) and Ref.~\cite{DiMauro:2019frs} (extending to 10 deg). Likewise the dashed lines are independent calculations for the MW. As can be seen, there is good consistency between the different estimates. Our resulting models are summarized in Table~\ref{tab:J_and_Cross}. 

We note that Ref.~\cite{lisanti2018search} reports an M31 $J$-factor (integrated within the scale radius) of $(6.2^{+7.9}_{-3.5})\times10^{19} \ \rm{GeV^2 \ cm^{-5}}$, corresponding to a boost factor of 2.64 and a scale radius of $2.57^\circ$. The uncertainty in their calculation comes from the uncertainty in $M_{\rm{vir}}$ and $c_{\rm{vir}}$. Their boost factor is comparable to our low and mid models (with an NFW profile). When integrating over the same scale radius, we obtain $J$-factor values in the range $2.2\times10^{19}$ $-$ $17.0\times10^{19}$ $\rm{GeV^2 \ cm^{-5}}$, in agreement with the values reported in Ref.~\cite{lisanti2018search}. 

\section{Results}
\label{sec:DM_parameterspace}

\begin{figure*}[t]
\centering
\includegraphics[width=0.75\textwidth]{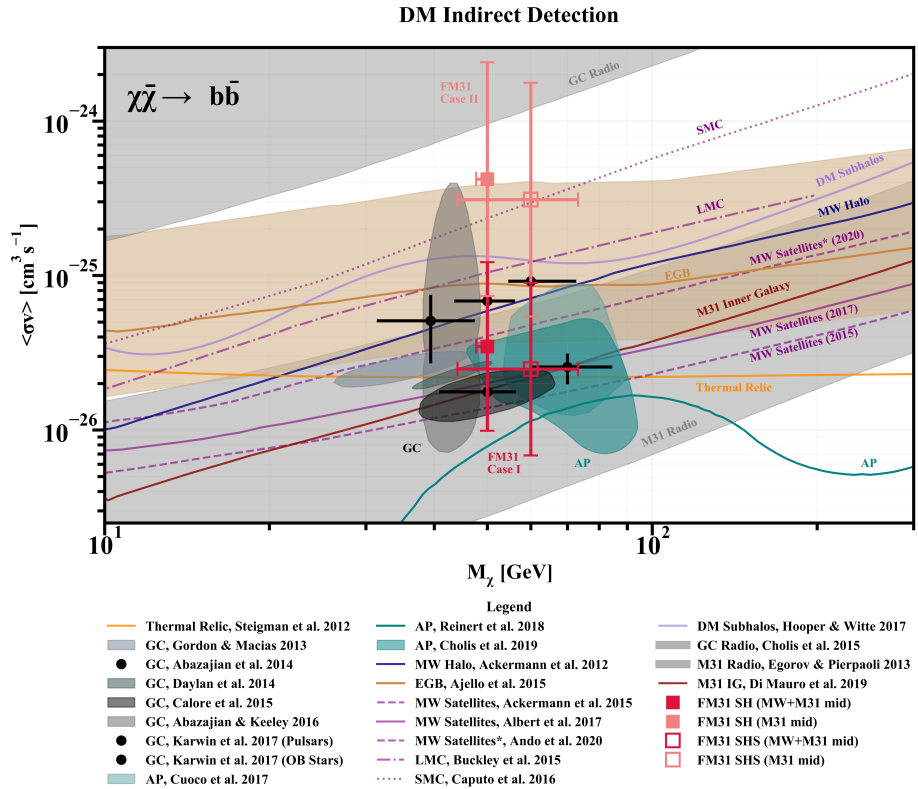}
\caption{DM parameter space. The red and coral data points are for M31's outer halo. The red data points correspond to case I, for which $J = \  J_{\mathrm{MW}} \ + \  J_{\mathrm{M31}}$. The coral data points are for case 2, for which $J =J_{\mathrm{M31}}$. The filled squares are for the SH fit, and the open squares are for the SHS fit. The results for the SHN are very similar to the SH, and so we don't include them here. Note that the error bars in the cross-section assume that the minimum subhalo mass is $10^{-6} \ M_\odot$, and they include the uncertainty due to the halo geometry. Contours for the GC excess are shown in black, and contours for the antiproton excess are shown in teal. Numerous limits from other targets are also overlaid, including the MW satellites shown with purple curves, and M31's inner galaxy shown with a red curve. See Section~\ref{sec:DM_parameterspace} for more details, as well as Appendix~\ref{sec:DM_parameter_space}.}
\label{fig:summary} 
\end{figure*}

We calculate annihilation cross-sections using Eq. \eqref{eq2} with the values obtained from following the procedure described in Sec.~\ref{sec:analysis}, and results are given in Table~\ref{tab:J_and_Cross}. There are two values for each cross-section; the first value results from fitting to the SH data (SHN data is very similar), and the second value results from fitting to the SHS data. In Figure~\ref{fig:summary} we plot the corresponding best-fit DM parameters. The red data points correspond to case I, for which $J =   J_{\mathrm{MW}} +  J_{\mathrm{M31}}$. The coral data points are for case II, for which $J =J_{\mathrm{M31}}$. The filled squares are for the SH fit, and the open squares are for the SHS fit. The results for the SHN fit are very similar to the SH, and so we do not include them here. Note that the error bars in the cross-section assume that the minimum subhalo mass is $10^{-6} \ M_\odot$, and they include the uncertainty due to the halo geometry outlined in Sec.~\ref{sec:analysis}. We compare the data points from M31's outer halo to numerous complementary targets for indirect DM searches. Details for all of the overlays are given in Appendix~\ref{sec:DM_parameter_space}.  

Broadly speaking, contours for the GC excess are shown in black, and contours for the antiproton excess are shown in teal. As can be seen, there is a rather large range in the different determinations. This is due to the different assumptions that are made in each analysis. Generally speaking, these results can be interpreted collectively as defining the currently explored systematic uncertainties in the respective signals. In the case of the GC excess, the uncertainty range in the cross-section spans roughly 1.5 orders of magnitude. This is because the GC excess is only a small fraction of the total emission in the region, and thus it has a strong dependence on the treatment of the IEM, which in general is difficult to accurately model due to the complexity of the GC region. Moreover, the inferred DM parameters also have a strong dependence on the halo assumptions, such as the local DM density, which may span between $\sim$0.3-0.6 $\mathrm{GeV/cm^{3}}$~\cite{Abazajian:2015raa,Evans:2018bqy}. In the case of the antiproton excess, Refs.~\cite{2017PhRvL.118s1102C,2019PhRvD..99j3026C} report detection contours, whereas Ref.~\cite{reinert2018precision} takes a less optimistic view, reporting upper limits (although the limits still clearly show an anomaly around the signal region).   

Another important constraint is the upper limits from the MW dwarfs. Here too there is a fairly large uncertainty range. Compared to the limits reported in Ref.~\cite{Ackermann:2015zua}, the latest limits from Ref.~\cite{Fermi-LAT:2016uux} are less constraining. These limits of course have a strong dependence on the assumptions made for the $J$-factors, and by employing semi-analytic models of DM subhalos to derive realistic satellite priors on the $J$-factor (for the ultrafaint dwarfs), Ref.~\cite{2020arXiv200211956A} has recently shown that the limits may be even weaker, by a factor of $\sim$2--7. Correspondingly, if the halos are non-spherical then the limits may be weakened as well, as discussed in Refs.~\cite{Bonnivard:2014kza,PhysRevD.95.123012}. 

As can be seen in Figure~\ref{fig:summary}, the limits coming from M31's inner galaxy are competitive with the limits from the MW dwarfs. In this case, however, the difficulty is in accurately separating a DM signal from the standard astrophysical emission. The limits shown in Figure~\ref{fig:summary} are from Ref.~\cite{DiMauro:2019frs}, and they are for the most conservative case, i.e.~they assume that all of the observed emission is from standard astrophysical processes, and thus model it using a $0.4^\circ$ disk, as determined from the emission itself. Upper limits for a DM signal are then calculated in addition to the disk. While this is definetly a very conservative choice to make, it is by no means preferred, as the $\gamma$-ray emission from M31's inner galaxy has actually been found to not correlate with regions rich in gas and star formation. 

The data points for M31's outer halo have a large overlap with the DM interpretations of both the GC excess and the antiproton excess, while also being compatible with the limits from the MW dwarfs. However, this requires that the $J$-factor be towards the higher end of the uncertainty range. Correspondingly, this has two main implications. First, the minimum subhalo mass must be $\lesssim 10^{-6} \ M_\odot$. Second, the signal must have some contribution from the MW's DM halo along the line of sight, i.e.~the $J$-factor must correspond to case I, as it cannot be due to M31 alone.  

\section{Summary, Discussion, and Conclusion}
\label{sec:conclusion}

An excess $\gamma$-ray signal towards the outer halo of M31 has recently been reported~\cite{karwin2019fermi}. In this work we interpret the excess in the framework of DM annihilation. As our representative case we use WIMP DM annihilating to bottom quarks, and we fit the DM mass and annihilation cross-section to the observed $\gamma$-ray spectra from Ref.~\cite{karwin2019fermi}. In that study M31's halo is characterized using three symmetric components centered at M31, namely, the IG (r $\leq 0.4^\circ$), SH ($0.4^\circ >$ r $\leq 8.5^\circ$), and FOH (r $> 8.5^\circ$). Here we fit just to the SH component. The IG and FOH components are difficult to disentangle from standard astrophysical processes and are not considered in this study.

The greatest uncertainty in our analysis is the determination of the $J$-factor, which we calculate using the CLUMPY code. This uncertainty arises from two main factors. First, there is a high uncertainty in the substructure nature of the DM halo's for both M31 and the MW, as well as an uncertainty in the halo geometries. To bracket the substructure uncertainty we vary the subhalo mass function, the fraction of the halo resolved in substructure, and the minimum subhalo mass in the ranges $1.9-2.0$, $0.12-0.35$, and $10^6 - 10^{-6} \ M_\odot$, respectively. For the concentration-mass relation we adopt the model from Ref.~\cite{Bullock:1999he}. The largest subhalo mass is taken to be 10\% the mass of the host halo. The calculations include 2 levels of substructure. For the underlying smooth density profiles we test both an NFW profile and an Einasto profile. The spatial distribution of subhaloes and the density profile of the subhaloes are assumed to be the same as the density profile of the main halo. All calculations are made self-consistently for M31 and the MW (i.e.~they have the same halo paramaters). Our calculated total boost factor ranges from $\sim$1.5$-$26.0 (for an NFW density profile). Note that this is the value reported by CLUMPY for the total halo, which we report here for easy comparison with other studies.

We have also characterized how the halo geometry impacts the $J$-factor for the M31 field. To do this we have used the range of different halo shapes found in the literature. For the MW we find that the halo shape may change the $J$-factor in the range $J/J_{\mathrm{Sph}} = 0.34-2.3$. The corresponding range for M31 is found to be 0.8--1.3. Thus the impact is more significant for the MW, due to our position within the halo. 

The other main uncertainty in the $J$-factor for the M31 field is the contribution from the MW's DM halo along the line of sight. In Ref.~\cite{karwin2019fermi} a detailed modeling of the foreground emission was performed, as well as an in-depth analysis of the corresponding systematic uncertainties. However, the model does not explicitly account for a potential contribution from the MW's extended DM halo. It is likely that such a signal could be (partially) absorbed by the isotropic component. The magnitude of this effect, however, depends on the specific halo geometry and substructure properties of the MW DM halo in the M31 field, which  are  not well constrained. In order to help control this, Ref.~\cite{karwin2019fermi} used a region below the M31 field to tune the isotropic normalization. Here, we improve on this determination by considering variations of the MW DM component in the M31 field and in the tuning region due to different  halo geometries. We find that the ratio is significant and, more specifically, in the range of $J_{\mathrm{MW}}/J{_\mathrm{TR}} = 0.8-1.4$. Thus even in the ideal case where the isotropic component is able to perfectly absorb the emission from the MW's DM halo, there could still be a gradient in the M31 field that is not included in the foreground model and is likely to be a significant component in this region. Since the uncertainty in the $J$-factor due to the contribution from the MW's DM halo along the line of sight is significant but cannot be precisely constrained, here we consider the two extreme cases: one where  none of the MW halo component has been absorbed by the isotropic component, and so $J_{\mathrm{total}} = J_{\mathrm{M31}} + J_{\mathrm{MW}}$ (case I); the other where  the MW component has been completely absorbed so that $J_{\mathrm{total}} = J_{\mathrm{M31}}$ (case II).

When these uncertainties are taken into account, we find that the observed excess in the outer halo of M31 favors a DM particle with a mass of $\sim$46--73 GeV. The full systematic uncertainty in the cross-section currently spans 2.5 orders of magnitude, ranging from $\sim8 \times 10^{-27}-4 \times 10^{-24} \ \mathrm{cm^3 \ s^{-1}}$. We compare  the best-fit DM parameters for M31's outer halo to numerous complementary targets. We conclude that for the DM interpretation of the M31 outer halo excess to be compatible with the GC excess, anti-proton excess, and current indirect detection constraints, it requires the $J$-factor to be towards the higher end of the uncertainty range. This in turn has two main implications. First, the minimum subhalo mass must be $\lesssim 10^{-6} \ M_\odot$. And in fact this is expected in the standard DM paradigm ($\Lambda$CDM). Second, the signal must have a significant contribution from the MW's DM halo along the line of sight, i.e.~it is too bright to be originating from M31 alone. This condition cannot be ruled out, and it is in fact likely that some fraction of the MW DM halo emission is embedded in the signal toward M31. This is a feature of the methodology employed to tune the MW foreground, as discussed in this paper. Given these conditions hold, we find that there is a large overlap with the DM interpretations of both the GC excess and the antiproton excess, while also being compatible with the limits from the MW dwarfs. Although the uncertainty in the current measurements is clearly far too large to make any robust conclusions (either positive or negative), this region in parameter space still remains viable for discovery of the DM particle. 

Future prospects to confirm the excess toward the outer halo of M31, and to better understand its nature, crucially rely on improvements in modeling the interstellar emission towards M31. Furthermore,  observations of the halos of other galaxies, e.g. M33, could provide a confirmation of this type of signal, provided sufficient data is available since the signal is predicted to be fainter there. Other prospects may include a study of the distribution of properties of the isotropic background around the direction to M31 and further out with a goal to see the distortions in the MW DM halo. Alternatively, constraints on the subhalo population by other astrophysical probes and, in turn, on their   contribution to the M31  signal, might  also provide a further test of the viability of the DM interpretation. 

\section*{Acknowledgements}
C.K.~is pleased to acknowledge conversations with Daniel McKeown and James Bullock. I.M.~acknowledges support from NASA grant No.~NNX17AB48G.

\hypersetup{urlcolor=black}
\bibliography{citations_master.bib}

\appendix

\section{DM Parameter Space}
\label{sec:DM_parameter_space}

Here we summarize all of the results overlaid in Figure~\ref{fig:summary}. The black data points (furthest four to the right) are for a DM interpretation of the GC excess, as presented in Ref.~\cite{Karwin:2016tsw}. The two points at lower energy are for two of the models employed for the fore/background $\gamma$-ray emission from the MW, \textit{OB stars index-scaled}, and the points at higher energy are for the other two models, \textit{pulsars index-scaled}. The NFW profile has $\gamma=1.0$ (upper) and $\gamma=1.2$ (lower). In addition, the NFW profile has $\rm{R_s}=20$ kpc and $\rho_\odot = 0.4 \ \rm{GeV\ cm^{-3}}$. Note that the annihilation final state preferred by the fit to the data favors mostly bottom-type quarks in the final state, with a small fraction of leptonic final states. Thus this model is not directly comparable to the other overlays which generally assume annihilation into a single final state. 

The black contour that is highly elongated in the y-direction is for the GC excess from Ref.~\cite{Abazajian:2015raa}. The contour represents the total uncertainty ($3\sigma$ statistical + systematic). The uncertainty is dominated by the systematics, and in particular, the value of  the local DM density (this study also considers uncertainties due to the index and scale radius of the DM profile, $\gamma$ and $R_s$). The upper region of the contour corresponds to $\rho_\odot = 0.28 \ \rm{GeV\ cm^{-3}}$ (which is taken as the benchmark value), and the lower region of the contour corresponds to $\rho_\odot = 0.49 \ \rm{GeV\ cm^{-3}}$. The shift occurs at a cross section value of $\sim6 \times 10^{-26} \ \rm{cm^3 \ s^{-1}}$. See Ref.~\cite{Abazajian:2015raa} for details. Also plotted in Figure~\ref{fig:summary} is the best-fit point from Ref.~\cite{Abazajian:2014fta} (the black data point to the far left). 

Other contours for the GC excess are also shown with different shades of grey. The lowest and darkest contour (2$\sigma$) is from Ref.~\cite{Calore:2014xka}, then above that is the contour (2$\sigma$) from Ref.~\cite{Daylan:2014rsa}, and above that is the contour from Ref.~\cite{Gordon:2013vta}. The NFW profiles for all of these contours have $\gamma=1.2$, $R_s=20$ kpc, and $\rho_\odot = 0.4 \ \rm{GeV\ cm^{-3}}$. 

The two lowest purple curves show limits for the MW satellite galaxies. The dashed curve is from Ref.~\cite{Ackermann:2015zua} and results from the combined analysis of 15 dwarf spheroidal galaxies using Pass-8 data. The solid curve is from Ref.~\cite{Fermi-LAT:2016uux} and results from the combined analysis of 45 stellar systems, including 28 kinematically confirmed dark-matter-dominated dwarf spheroidal galaxies, and 17 recently discovered systems that are dwarf candidates. Note that the dwarf limits are obtained by assuming spherical symmetry of the DM halos; however, if the halos are non-spherical then the limits may be weakened, as discussed in Refs.~\cite{Bonnivard:2014kza,PhysRevD.95.123012}. We also plot the limits from Ref.~\cite{2020arXiv200211956A} ($V_{50} = 10.5 \mathrm{\ km \ s^{-1}}$), which employs semi-analytic models of DM subhalos to derive realistic satellite priors on the $J$-factor (for the ultrafaint dwarfs). This result explicitly exemplifies the uncertainty range associated with limits from the MW dwarfs. 

The two highest purple curves are for the LMC and SMC. The dash-dot curve shows $2\sigma$ limits from the LMC from Ref.~\cite{Buckley:2015doa}, based on Pass-7 data. The dotted curve shows $2\sigma$ limits from the SMC from Ref.~\cite{Caputo:2016ryl}. 

The tan band shows the 2$\sigma$ upper-limit from the extragalactic $\gamma$-ray background (EGB) from Ref.~\cite{Ajello:2015mfa}. The band reflects the uncertainties related to the modeling of DM subhaloes. This analysis shows that blazars, star-forming galaxies, and radio galaxies can naturally account for the amplitude and spectral shape of the EGB over the energy range 0.1--820 GeV, leaving only modest room for other contributions.   

The blue curve shows $\gamma$-ray limits (3$\sigma$) from the MW halo from Ref.~\cite{Ackermann:2012rg}. This is the limit obtained with modeling the MW diffuse emission using GALPROP, for an NFW profile, with $\gamma=1$ and a local DM density of 0.43 GeV $\rm{cm^{-3}}$. The limits are generally weaker without modeling the diffuse emission, and they have a strong dependence on the local DM density.

The light purple curve is for DM subhaloes from Ref.~\cite{hooper2017gamma}. These limits are based on DM subhalo candidates from the unassociated point sources detected by \textit{Fermi}-LAT. In total there are 19 subhalo candidates. The minimum subhalo mass for the upper limit calculation is assumed to be $\rm{10^{-5} \ M_\odot}$.

The upper gray band  in Figure~\ref{fig:summary} shows radio constraints for the GC from Ref.~\cite{2015PhRvD..91h3507C}. The limits are derived using VLA observations at 330 MHz of the central $0.04^\circ$ around Sgr A*. An NFW profile is used with $\gamma=1.26$, $\rm{R_s=20}$ kpc, a local DM density of 0.3 GeV $\rm{cm^{-3}}$, and a flat density core of 2 pc. The limits include energy losses due to IC and convection. The lower limit is for $V_C=0 \ \rm{km \ s^{-1}}$, and the upper limit (not shown) is for $V_C=1000 \ \rm{km \ s^{-1}}$. The limits can be much stronger (up to 3 or 4 orders of magnitude) when not including IC and convection, or for a core radius closer to zero. There is also a high uncertainty of the magnetic field strength in the innermost region of the GC.

The lower gray band shows radio limits from the central region of M31 ($\sim$1 kpc) from Ref.~\cite{Egorov:2013exa}. The band represents joint constraint from four different surveys: VLSS (74 MHz), WENSS (325 MHz), NVSS (1400 MHz), and GB6 (4850 MHz). An M31 signal is detected for all surveys but VLSS. The highest region is for a central magnetic field strength $B_0 =5$ $\mu$G and DM concentration of $c_{100}=12$, the middle region is for $B_0 =50$ $\mu$G and DM concentration of $c_{100}=20$, and the lowest region is for  $B_0 =300$ $\mu$G and DM concentration of $c_{100}=28$. An NFW profile is used for the DM density, with $\gamma=1$, and a flat core for r$<$50 pc. The limits have a large uncertainty due to the uncertainties in the DM profile and magnetic field strength in the inner regions of M31. The magnetic field is modeled with an exponential dependence in galactocentric radius and height above the galactic plane. The analysis accounts for leptonic energy losses due to IC emission, synchrotron emission, Bremsstrahlung, and Coulomb scattering, with synchrotron emission being the dominant loss mechanism over most of the energy range. We note, however, that uncertainties in the astrophysical modeling of these processes may weaken the limits even further. In particular, the limits have a strong dependence on the relative strength of the inverse Compton losses compared to the synchrotron losses, which in turn depends on the energy density of M31's interstellar radiation field. 

Also shown are contours for a recently reported excesses in the flux of antiprotons. The upper light teal contour (2$\sigma$) is from Ref.~\cite{2017PhRvL.118s1102C}. The lower dark contour (2$\sigma$) is from Ref.~\cite{2019PhRvD..99j3026C}. The NFW profiles for these contours have $\gamma=1.0$, $R_s=20$ kpc, and $\rho_\odot = 0.4 \ \rm{GeV\ cm^{-3}}$. The teal curve shows upper-limits from Ref.~\cite{reinert2018precision}, where a less optimistic view of the excess is given (although the limits still clearly show an anomaly around the signal region).
 
The red curve is for M31's inner galaxy from Ref.~\cite{DiMauro:2019frs}. These limits are obtained by assuming that all of  the observed $\gamma$-ray emission from M31's inner galaxy arises from standard astrophysical emission, and therefore including a $0.4^\circ$ disk template (which is derived directly from the bright $\gamma$-ray emission that is observed) in the DM fit. In addition, to account for the foreground/background emission, the standard IEM is fit directly to the $\gamma$-ray data in the signal region.

\end{document}